\documentclass[sigconf]{acmart}
\AtBeginDocument{%
 \providecommand\BibTeX{{%
 \normalfont B\kern-0.5em{\scshape i\kern-0.25em b}\kern-0.8em\TeX}}}

\copyrightyear{2024}
\acmYear{2024}
\setcopyright{acmlicensed}
\acmConference[CHI '24]{Proceedings of the CHI Conference on Human Factors in Computing Systems}{May 11--16, 2024}{Honolulu, HI, USA}
\acmBooktitle{Proceedings of the CHI Conference on Human Factors in Computing Systems (CHI '24), May 11--16, 2024, Honolulu, HI, USA}
\acmDOI{10.1145/3613904.3642317}
\acmISBN{979-8-4007-0330-0/24/05}

\usepackage{multirow}
\usepackage{subcaption}
\usepackage{acmart-taps}
\usepackage{balance}       
\usepackage{graphics}      
\usepackage{hyperref}
\usepackage{color}
\usepackage{textcomp}
\usepackage{subcaption}
\usepackage{xcolor}
\usepackage{makecell}
\usepackage{multirow}
\usepackage{multicol}
\usepackage{enumitem}
\usepackage{amsmath}
\usepackage{stfloats}
\usepackage{graphicx}
\usepackage{amsthm}
\usepackage{listings}
\usepackage{xspace}
\usepackage[linesnumbered]{algorithm2e}
\usepackage{tabularx}
\usepackage{arydshln}

\newcolumntype{L}[1]{>{\raggedright\let\newline\\\arraybackslash\hspace{0pt}}m{#1}}
\newcolumntype{C}[1]{>{\centering\let\newline\\\arraybackslash\hspace{0pt}}m{#1}}
\newcolumntype{R}[1]{>{\raggedleft\let\newline\\\arraybackslash\hspace{0pt}}m{#1}}

\makeatletter
\def\thickhline{%
  \noalign{\ifnum0=`}\fi\hrule \@height \thickarrayrulewidth \futurelet
   \reserved@a\@xthickhline}
\def\@xthickhline{\ifx\reserved@a\thickhline
               \vskip\doublerulesep
               \vskip-\thickarrayrulewidth
             \fi
      \ifnum0=`{\fi}}
\makeatother

\makeatletter
\def\thickhlinespace{%
  \addlinespace[1ex]
  \noalign{\ifnum0=`}\fi\hrule \@height \thickarrayrulewidth \futurelet
   \reserved@a\@xthickhline
   \addlinespace[1ex]
   }
\def\@xthickhlinespace{\ifx\reserved@a\thickhline
               \vskip\doublerulesep
               \vskip-\thickarrayrulewidth
             \fi
      \ifnum0=`{\fi}}
\makeatother

\newlength{\thickarrayrulewidth}
\newlength\Origarrayrulewidth

\newenvironment{s_itemize}{
\begin{itemize}
  \setlength{\itemsep}{3pt}
  \setlength{\parskip}{0pt}
  \setlength{\parsep}{0pt}
}{\end{itemize}}

\definecolor{downredcolor}{HTML}{e31a1c}
\definecolor{upgreencolor}{HTML}{33a02c}

\definecolor{DarkGreen}{HTML}{5DAC81}

\begin{document}

\title{InteractOut: Leveraging Interaction Proxies as Input Manipulation Strategies for Reducing Smartphone Overuse}

\author{Tao Lu}
\affiliation{%
  \institution{University of Michigan}
  \city{Ann Arbor, MI}
  \country{USA}}
\email{luttul@umich.edu}

\author{Hongxiao Zheng}
\affiliation{%
  \institution{University of Michigan}
  \city{Ann Arbor, MI}
  \country{USA}
}
\email{waleyz@umich.edu}

\author{Tianying Zhang}
\affiliation{%
  \institution{University of Michigan}
  \city{Ann Arbor, MI}
  \country{USA}
}
\email{danniez@umich.edu}

\author{Xuhai Xu}
\affiliation{%
  \institution{Massachusetts Institute of Technology}
  \city{Cambridge, MA}
  \country{USA}
}
\email{xoxu@mit.edu}

\author{Anhong Guo}
\affiliation{%
  \institution{University of Michigan}
  \city{Ann Arbor, MI}
  \country{USA}}
\email{anhong@umich.edu}

\renewcommand{\shortauthors}{Lu et al.}

\begin{abstract}
Smartphone overuse poses risks to people's physical and mental health. However, current intervention techniques mainly focus on explicitly changing screen content (i.e., output) and often fail to persistently reduce smartphone overuse due to being over-restrictive or over-flexible. We present the design and implementation of InteractOut, a suite of implicit input manipulation techniques that leverage interaction proxies to weakly inhibit the natural execution of common user gestures on mobile devices. We present a design space for input manipulations and demonstrate 8 Android implementations of input interventions. We first conducted a pilot lab study (N=30) to evaluate the usability of these interventions. Based on the results, we then performed a 5-week within-subject field experiment (N=42) to evaluate InteractOut in real-world scenarios. Compared to the traditional and common timed lockout technique, InteractOut significantly reduced the usage time by an additional 15.6\% and opening frequency by 16.5\% on participant-selected target apps. InteractOut also achieved a 25.3\% higher user acceptance rate, and resulted in less frustration and better user experience according to participants' subjective feedback. InteractOut demonstrates a new direction for smartphone overuse intervention and serves as a strong complementary set of techniques with existing methods.
\end{abstract}



\settopmatter{printfolios=true}

\maketitle

\section{Introduction}

The rapid growth of digital technologies in the past two decades has led to a boom in convenient access for users to digital products and services. Mobile devices are one of the most highly used digital hardware technologies~\cite{CV19DigTech}, which have become even more prevalent during the COVID-19 \cite{CV19Screentime, CV19DigTech}. Recent research has revealed concerns about the overuse of digital devices, such as poor sleep, impaired eye health, digital device addiction, and academic distraction \cite{ScreentimeChildEffects, CV19eyes, CV19Myopia, SmartphoneDependence, 24HrNoNotifications, MessageDistraction, FacebookImpactGrades}. Many users are aware of these concerns~\cite{lukoff2018makes}, and there is a growing need to reduce mobile device usage~\cite{lyngs2019self}.

In response, there has been a wide range of research on smartphone overuse intervention techniques. Prior studies mainly focus on two directions: \textit{(i)} restrictive blocking that introduces interaction barriers to encourage users to stop smartphone entertainment (e.g., \cite{GoalKeeper, InteractionRestraint, LocknType, TypeOut}), and \textit{(ii)} alerts or reminders that direct users away from smartphones and penalize overuse (e.g., \cite{MyTime, NUGU, Goodvibrations, GoldenTime}). However, blocking often becomes over-restrictive and causes a bad user experience, and some can even backfire and cause users to have more smartphone usage~\cite{GoalKeeper, TypeOut}. On the other hand, reminders tend to be over-flexible and can be easily ignored. Therefore, there is a need for new intervention techniques that strike the right balance between restrictiveness and flexibility.
Our work aims to achieve this goal.

Most of the prior intervention techniques adopted an explicit pattern to change the smartphone's \textit{output} (e.g., locking, showing a notification).
Many of these techniques are based on the Dual Process Theory~\cite{hofmann_impulse_2009}, where changing the output can potentially prevent System 1 (an intuitional and unconscious process, in this case, the desire for immediate gratification of using smartphones) and trigger System 2 (a logical and analytical process, in this case, the reasoning and reflection on whether they should continue smartphone usage)~\cite{TypeOut,lyngs2019self}.
However, there is very little work exploring the implicit intervention pattern that alters users' \textit{input} to the smartphone. Prior work in cognitive science and psychology~\cite{alter2007overcoming,evans2013dual} suggests that the mismatch between the expected results (based on the issued command, i.e., user input) and the actual outcome (after input manipulation~\cite{norman2010gestural}) can provide an opportunity to trigger analytic reasoning, i.e., System 2.
This suggests a promising direction for smartphone overuse intervention.

We present \textit{InteractOut}, a suite of implicit input manipulation techniques that slightly inhibit the natural execution of common user gestures on smartphones, such as taps and swipes. 
These input manipulation techniques introduce interaction costs and decrease the smoothness of smartphone interaction to nudge users towards reducing usage. By varying the intervention intensity, we can have precise control over the balance between restrictiveness and flexibility.
We present a design space of continuous and discrete manipulations on time, location, direction, and the number of fingers, which are the four main elements of gesture commands. Our manipulation strategies include multiple remapping actions, such as shift, complicate, extend, and disable (Figure \ref{fig:design-space-diagram}).
Inspired by Interaction Proxies \cite{InteractionProxy}, we then demonstrate 8 Android implementations of input interventions throughout our design space, including four tap interventions (Tap Delay, Tap Prolong, Tap Shift, and Tap Double), as well as four swipe interventions (Swipe Delay, Swipe Deceleration, Swipe Multiple Fingers, and Swipe Reverse).

We first conducted a pilot study in the lab (N=30) to evaluate how users respond to the interventions, which helped us fine-tune the specific parameters, and understand what might contribute to their effectiveness.
Based on the lab study, we identified four most promising interventions (Tap Delay, Tap Shift, Swipe Delay, and Swipe Deceleration) and their intensities. We then performed a five-week field experiment to measure their effectiveness and user receptivity in a real-world deployment. We deployed InteractOut with 42 participants in an uncontrolled everyday setting, where users can make their own decisions of smartphone usage in situ. We compared InteractOut against the timed lockout baseline, one of the most common intervention techniques supported by modern Android and iOS operating systems~\cite{android_screentimer,ios_screentimer}.
Our results show that InteractOut significantly outperformed the lockout baseline by reducing the usage time by an additional 15.0\% and reducing the opening frequency by an additional 17.0\% on target apps. Meanwhile, InteractOut achieved a 25.4\% higher user acceptance rate than lockout, which indicated that participants had a significantly higher receptivity towards our techniques.
Moreover, participants' subjective feedback indicated that they preferred the InteractOut interventions.
Users could still continue using their phones with some friction on input. But unlike the timed lockout, where users may forget the time limit they requested, InteractOut presents continuous reminders that are more acceptable, has both lower user-perceived friction and higher effectiveness, and strikes a good balance between restrictiveness and flexibility.

InteractOut introduces a new perspective of smartphone overuse interventions that can be compatible with and complementary to existing intervention techniques. The main contributions of our work include:
\vspace{-.5pc}
\begin{itemize}
\item We introduce InteractOut, a novel set of implicit input manipulation techniques for smartphone overuse intervention. We present a design space of input manipulations that weakly inhibits the natural execution of common user gestures.
\item We leverage interaction proxies and develop eight Android implementations of input intervention techniques throughout our design space to illustrate their expressiveness.
\item We conduct a pilot lab study (N=30) to evaluate the usability of different intervention techniques. Our results identify four most promising intervention techniques: Tap Delay, Tap Shift, Swipe Delay, and Swipe Deceleration.
\item We further conduct a five-week deployment study (N=42) in the wild. Our results demonstrate significant advantages of InteractOut over the common timed lockout baseline technique on smartphone usage time, opening frequency, user acceptance rate, and subjective user experience.
\end{itemize}

\section{Related Work}
We first summarize existing intervention techniques for smartphone overuse reduction. We then introduce the theoretical foundation of our input manipulation design. Finally, we introduce related work in interaction proxies, which inspired our intervention design.

\subsection{Designs for Smartphone Overuse Intervention}

In an age where smartphones have become an integral part of our lives, their overuse has become an increasingly important problem, posing risks for physical and mental health \cite{CV19Myopia, MentalHealthEffect, SmartphoneDependence, Nomophobia, ProbelmPsychosocial, SmartphoneBurnout}. There exist many research studies, and commercial technologies that offer resources to aid in the reduction of smartphone overuse \cite{InterventionFramework, AppDetox}. According to the strength of obstacles to limit smartphone usage \cite{CommericalAppComp}, current intervention techniques can be categorized into two groups: \textit{(i)} strong interventions with restrictive blocking, and \textit{(ii)} soft interventions with alerts or reminders.

Strong intervention techniques often block smartphone usage, forcing users to stop using the device. Such restrictive and coercive lockout was often used for productivity because it is the simplest way to control smartphone usage, although it often causes user frustration. For example, GoalKeeper \cite{GoalKeeper} studied users' receptivity to different lockout intensity levels and found that users were frustrated by restrictive and coercive interventions. Interaction Restraint \cite{InteractionRestraint} and Lock'n'Type \cite{LocknType} explored how cognitive tasks like typing a series of digits could help reduce users' violation of their predefined smartphone usage limit. TypeOut \cite{TypeOut} is a more recent study combining self-affirmation \cite{SelfAffirmationPsych} and just-in-time intervention \cite{JIT} to further reduce the annoyance of strong lockout interventions. In commercial applications, the lockout mechanism is also commonly used. For instance, Forest \cite{Forest}, PromodoLock \cite{PromodoLock}, and LessPhone \cite{LessPhone} are popular focusing tools using strong lockout to reduce smartphone distractions.

Soft intervention techniques, on the other hand, do not directly limit usage, and instead try to persuade users to reduce usage. Common examples include monitoring and notifications. For instance, MyTime \cite{MyTime} used persistent notifications that displayed usage to reduce users' desire to use smartphones. Social impact and competition have also been utilized to make smartphones less entertaining. NUGU \cite{NUGU} visualized smartphone usage on competitive ranking scoreboards to facilitate self-motivation. Lockn'LoL \cite{LocknLoL} also used social effects but focused on smaller groups of friends.
Let's FOCUS \cite{LetsFOCUS} specifically targeted the distraction problem in the classroom using both reminders and bounded space, limiting most smartphone functionalities.
In commercial products, using social competition is also common, such as AntiSocial \cite{AntiSocial}, ActionDash \cite{ActionDash}, and RealizD \cite{RealizD}.
Besides introducing contents of screen output,
Good Vibrations \cite{Goodvibrations} studied the effect of subtle, repeating phone vibrations on reducing smartphone overuse.

However, existing techniques are either overly restrictive (i.e., strong intervention) that cause user frustration, or are overly flexible (i.e., soft intervention) that lead to low engagement. For both types of interventions, the effectiveness of existing techniques often comes with an expense of disrupting users. This is often unacceptable when users are in the middle of an important task (e.g., sending an important email).
There is a need for new intervention techniques that strike the right balance.
Meanwhile, existing intervention techniques mostly adopted an explicit pattern to change the smartphone's \textit{output} (content on the screen, e.g., locking, showing a notification). However, there is no prior work exploring the implicit intervention pattern that alters users' \textit{input} for smartphone intervention.
Our work aims to strike the right balance between restrictiveness and flexibility by altering users' input.

\subsection{Cognitive Theory Foundation of Input Manipulations}
To design an effective intervention technique with minimal user frustration or annoyance, we need to investigate the underlying mental factors that lead to users' decisions to continue or stop smartphone usage.
This decision-making process can be largely explained by Dual Process Theory \cite{hofmann_impulse_2009, TypeOut, wu2024mindshift}. In this theory, human behaviors are mentally controlled by two ``systems'' of thinking: an intuitional, unconscious, and spontaneous process (System 1) and an analytical, self-aware, and deliberative process (System 2). To engage users, app designers often try to cater to one's desire for immediate gratification and override their logical thinking \cite{valasek2022disciplining, lim2017distraction, lyngs2019self}.
In smartphone applications such as social media platforms, designers use techniques (e.g., following and recommendation features) that constantly refresh bite-size content to attract user attention. To obtain this content, users often just need to use simple tap-and-swipe interactions~\cite{tor2021impact, hintze2017large, cho2021reflect}. Such low cognitive and dexterity requirements can lead to longer usage and eventually addiction to the application.

Reversely, researchers found that harder or more sophisticated interactions can evoke users' System 2 thinking and increase their awareness of smartphone usage \cite{tor2021impact}. Moreover, existing research in behavior science showed that the mismatch between the expected results and the actual outcome can provide an opportunity to trigger System 2 thinking~\cite{alter2007overcoming,taleb2014antifragile}. This sheds light on our idea of manipulating \textit{input} to introduce a mismatch between the issued command and the output on the screen~\cite{norman2010gestural}.
We implement this idea, explore the design space of input manipulations, and build InteractOut. More specifically, we modify users' tap and swipe gestures such that they do not receive their expected output change on the screen in time. This expectation mismatch weakens System 1, so that System 2 can take more control of users' decisions.

\subsection{Input Manipulation and Interaction Remapping}

There has been a wealth of research to improve user experience and accessibility of application interfaces through user interaction remapping~\cite{TheFatThumb, umaaresam2015, rico2010usable}.
For example, Sugilite \cite{SUGILITE} is a programming-by-demonstration system that enables users to map a set of UI interaction workflows into a single command. Commercial technologies such as iOS Shortcuts \cite{Shortcuts} and IFTTT \cite{IFTTT} also provide features that allow users to create personalized interaction remappings to simplify phone use. These techniques save operations for repetitive tasks by remapping a number of taps and swipes into zero or a few button clicks. 
Closer to our work, Interaction Proxies \cite{InteractionProxy} presented a framework to modify user interactions. It serves as a layer between an application's original interface and the manifest interface that a person uses to perceive and manipulate the application. This strategy allows third-party developers and researchers to modify the app interactions without modifying the original application.
The initial purpose of interaction proxies is to minimize the gulf of command execution \cite{Gulfs}, making applications easier to use and improving their accessibility. Their method of using a proxy layer to add or reinterpret the accessibility metadata demonstrates a simple way of runtime modification on smartphone applications.
We are inspired by this method and investigate how to use it in the ``opposite'' way: we can leverage interaction proxies to manipulate input and make phones ``harder to use.''
InteractOut utilizes interaction proxies to increase the difficulty for users to accomplish desired actions on mobile app interfaces.
We investigate the effects of such intervention design in smartphone overuse reduction.
\section{InteractOut}
\label{sec:system}
We first present the design space of implicit input manipulation techniques for overuse intervention. We then demonstrate how our intervention works on the Android operating system, along with 8 Android implementations to instantiate our design space.

\subsection{InteractOut Design Space}
\label{sub:system:design-space}
Most modern smartphones use a touch screen as the primary way for interaction. Touch gestures have become increasingly sophisticated, allowing users to manipulate their smartphones in more convenient and advanced ways \cite{TSGestureDesigns, ShortcutGesture}.
Still, simple tap and swipe gestures remain the most fundamental and common gestures due to their low cognitive cost \cite{neyman2017survey}. Thus, we categorize touch-based interactions into two major groups, tap-based and swipe-based, to facilitate a holistic and systematic intervention design.

\textit{Tap-Based Smartphone Interactions.}
Most interactions users perform on the smartphone are tap-based. This group of interactions does not have continuous moves on the screen but generally requires a precise location. The location is associated with the user's intention of the tap, such as selection, confirmation, and cancellation. The common types of taps include single tap, double tap, long press (fixed duration), and hold (indefinite duration). Our intervention design centers on single and double taps. We focus on manipulating their \textit{time} (i.e., duration) and \textit{location}.

\textit{Swipe-Based Smartphone Interactions.}
Swipe-based interactions are another group of common interactions. They have continuous and directed movement on the screen, and the trajectories are the most important information of a swipe. Swipe-based interactions are often used for navigation and adjustment. For example, users swipe a list to view hidden items due to limited screen sizes or pinch the picture to examine the details. For swipe-based interactions, we focus on the manipulation of \textit{time}, \textit{number of fingers}, and \textit{direction}.

\begin{figure*}[t!]
 \centering
 \includegraphics[width=.75\linewidth]{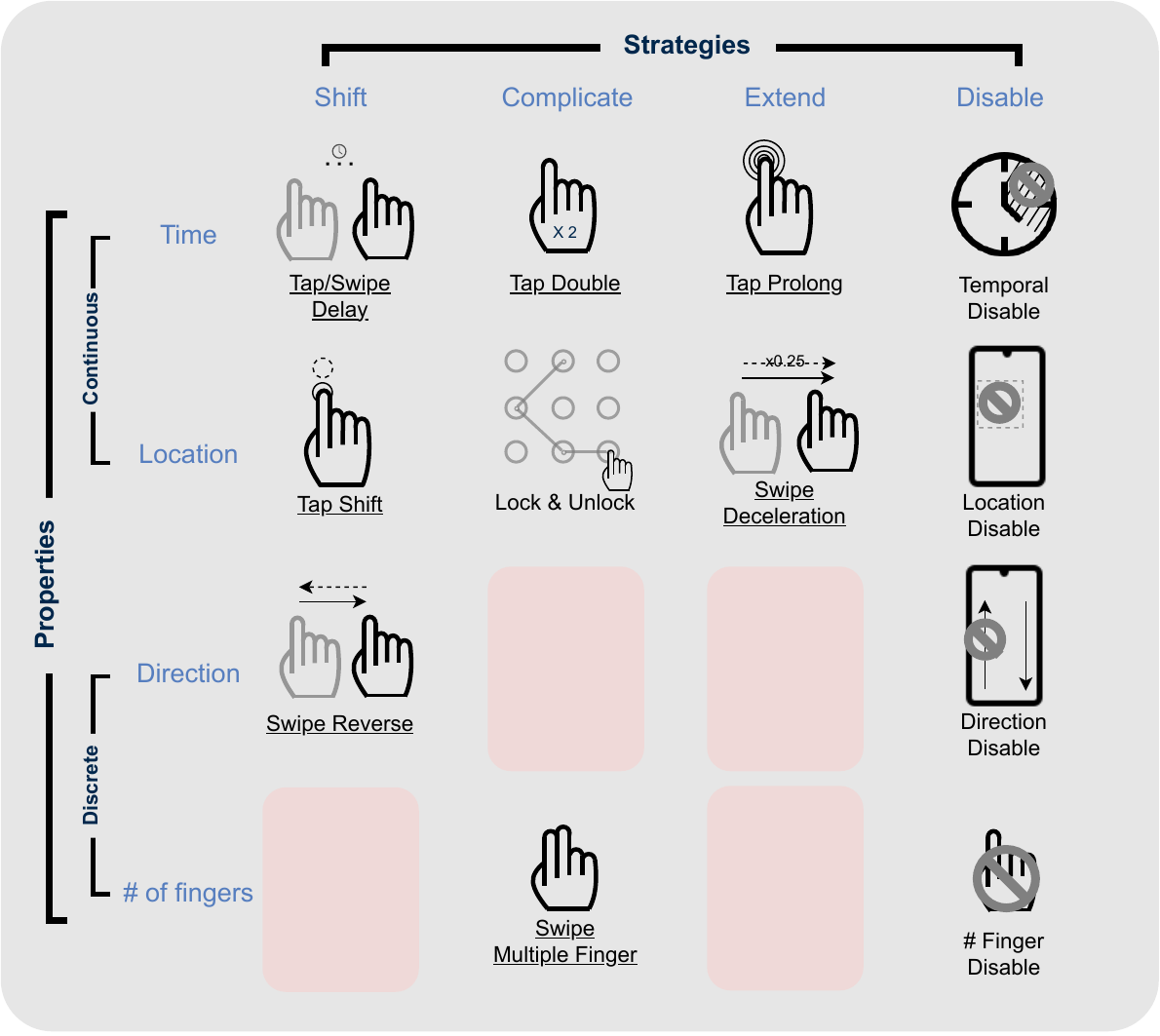}
 \vspace{-.5pc}
 \caption{Design Space of InteractOut. The rows include the four essential interaction properties, and the columns show the four manipulation strategies.
 The underlined techniques are our demonstrated implementations, while others are techniques associated with interactions. Note that pink blocks indicate invalid types.}
 \Description{Design Space of InteractOut. It manipulates the time, location, direction, and number of finger properties of gestures by four strategies: shift, complicate, extend, and disable.}
 \vspace{-1pc}
 \label{fig:design-space-diagram}
\end{figure*}

We present a design space for our smartphone interventions. Based on the categorization above, we focus on four essential properties of the interaction: (1) \textit{Time}, (2) \textit{Location}, (3) \textit{Direction}, and (4) \textit{Number of fingers}. To manipulate these properties, we propose four types of strategies that change input from different perspectives:

\begin{enumerate}[label=(\alph*)]
\item[(a)] \textit{Shift}: Changing the input outcome based on the interaction property. This can be applied to \textit{Time}, \textit{Location}, and \textit{Direction}.
\item[(b)] \textit{Complicate}: Asking for more operations from users to achieve the expected effect. This can be applied to \textit{Time}, \textit{Location}, and \textit{Number of fingers}.
\item[(c)] \textit{Extend}: Requiring extended and longer interaction from users to achieve the expected effect. This can be applied to \textit{Time} and \textit{Location}.
\item[(d)] \textit{Disable}: Nullifying the input effect of the property. This can be applied to all four properties, including \textit{Time}, \textit{Location}, \textit{Direction}, and \textit{Number of fingers}.
\end{enumerate}

This leads to a 4$\times$4 matrix for potential intervention types, as shown in Figure \ref{fig:design-space-diagram}.
After removing invalid ones (shown as pink blocks in Figure~\ref{fig:design-space-diagram}), our design space includes 12 intervention types, including a few popular existing techniques, such as Lock \& Unlock~\cite{ios_screentimer, android_screentimer} and the Temporal Disable strategy~\cite{Forest, GoalKeeper}.
Here we propose 8 input manipulation intervention techniques.

\begin{itemize}
 \item Tap Delay: Users' tap is postponed for a certain period before it takes effect;
 \item Tap Prolong: Users need to press the screen longer than a threshold to trigger a tap; 
 \item Tap Shift: Users' tap is shifted in a fixed direction away from the actual tap position;
 \item Tap Double: Users need to tap twice to trigger a single tap;
 \item Swipe Delay: Users' swipe is postponed for a certain period before it takes effect;
 \item Swipe Deceleration: The effect of a swipe is slower than the users' actual gesture trajectory;
 \item Swipe Reverse: Users' swipe direction is reversed;
 \item Swipe Multiple Fingers: Users need to use more than one finger to swipe.
\end{itemize}

\begin{figure*}[t!]
 \centering
 \includegraphics[width=\linewidth]{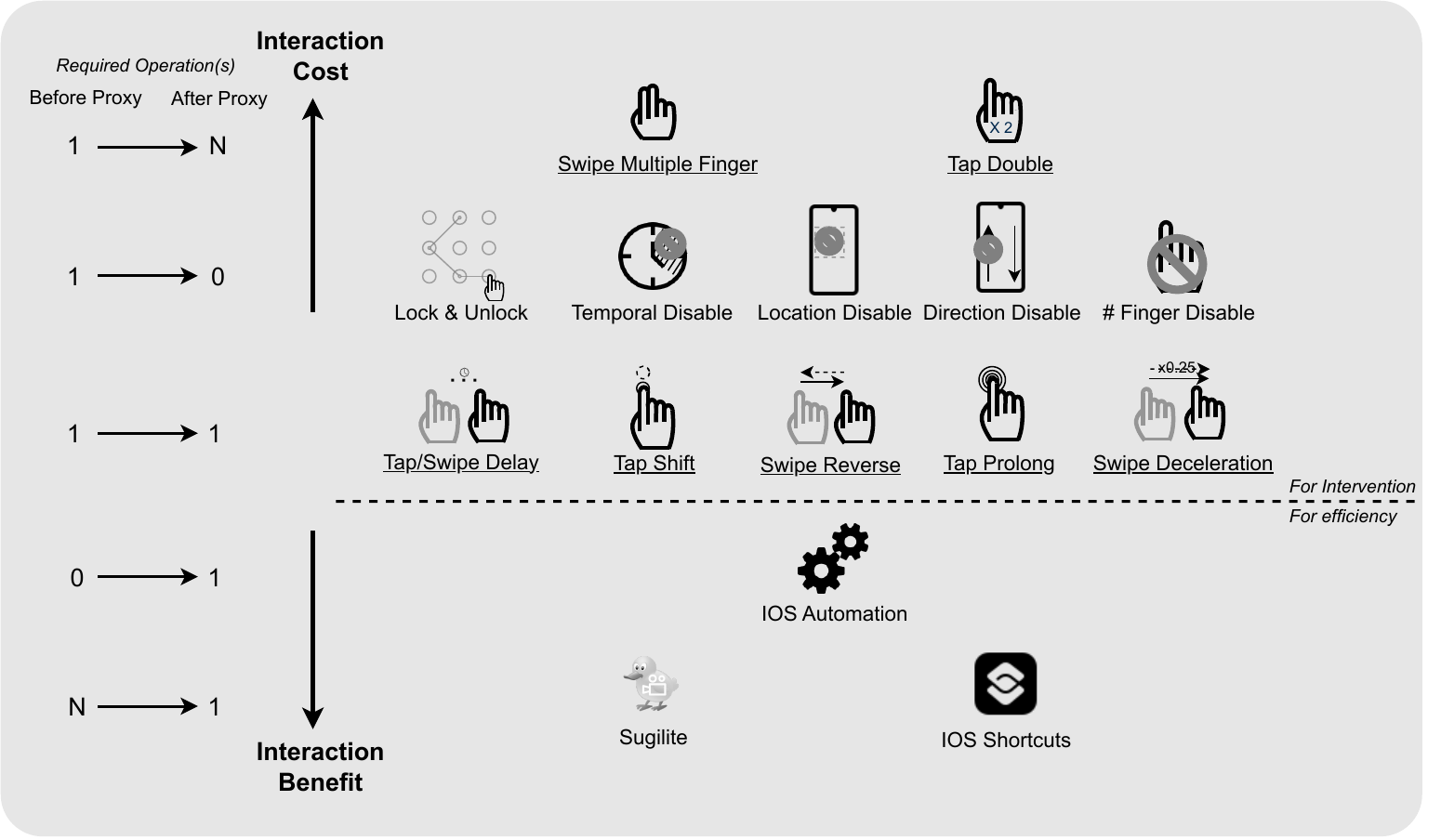}
 \vspace{-1.5pc}
 \caption{Summary of Remapping Methods According to the Design Space of Interaction Proxies \cite{InteractionProxy}. The underlined techniques are our demonstrated implementations. ``1'', ``0'', and ``N'' here means the number of operations/interactions needed to perform a task before and after the proxy.}
 \Description{InteractOut Remapping Framework. The blocks with solid borders are our demonstrated implementations. They all lie in the area of making things harder. The tools developed for efficiency and accessibility are making things easier.}
 \label{fig:remapping}
\end{figure*}

Leveraging Interaction Proxy remapping~\cite{InteractionProxy}, these intervention techniques replace the original interaction with more difficult and counter-intuitive micro-tasks that require more physical or mental efforts.
Figure \ref{fig:remapping} presents a summary of the remapping methods which includes our design space as well as existing remapping techniques. Note that most existing remapping techniques are designed to facilitate interaction rather than to introduce friction.

There are five remaining interventions our design space affords, including (1) Lock \& Unlock: Users need to do some tasks to continue the usage, i.e., sliding a block or typing numbers; (2) Temporal Disable: The screen is disabled for a period of time; (3) Location Disable \footnote{Note that there are also disabling techniques based on the geographical location of the smartphone~\cite{LetsFOCUS, Shortcuts}, and they are different from the ``Location Disable'' we present in the design space.}: Certain areas of the screen is disabled; (4) Direction Disable: Certain direction of interaction (mainly swipe) is disabled; and (5) \# of Finger Disable: The operation (mainly tap) is disabled if performed with a certain number of fingers.
Existing intervention techniques often combine the Temporal Disable and Lock \& Unlock to block smartphone usage. Users set a time limit and the app/screen will be blocked when the usage exceed the limit (i.e., Temporal Disable). Users have to stop or follow a specific unlock procedure to continue (i.e., Lock \& Unlock).
In this paper, we do not cover the four disable interventions as they are overly restrictive and would compromise user experience.
We use Lock \& Unlock as the baseline as it is already widely adopted by modern smartphones.

\subsection{InteractOut Mobile App}
\label{sub:mobile}

\paragraph*{Intervention Implementations}

\begin{figure*}[t!]
 \centering
 \includegraphics[width=\linewidth]{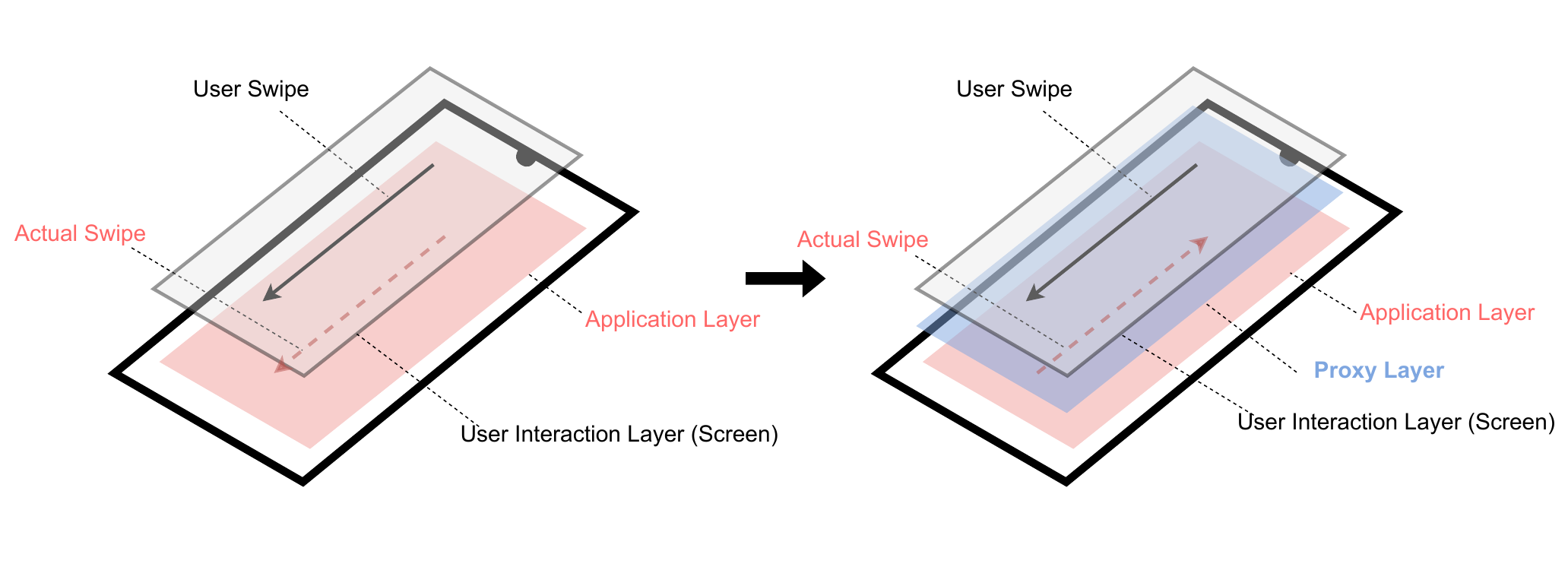}
 \vspace{-3pc}
 \caption{Structure of the InteractOut Implementation on Android. The proxy overlay takes user input and outputs the modified gesture. Swipe Reverse intervention is used for illustration.}
 \Description{Different from the original structure used in Interaction Proxy, The structure we use in InteractOut has only the proxy layer, which is an overlay in Android implementation. This layer has a Gesture Detector and is able to recognize the user's gestures. Then it can construct an artificial gesture and do some modifications if necessary. Finally, it outputs the constructed gesture, and users can use the interface changes.}
 \label{fig:overlay-structure}
 \vspace{-1pc}
\end{figure*}

We instantiate InteractOut as an application on the Android system.
The left of Figure \ref{fig:overlay-structure} shows the case when users operate the phone normally. Users perform gestures on the screen. The phone recognizes the operation and passes it to the application layer. The application responds according to the operation (e.g., display new content with a swipe down). 
InteractOut inserts a proxy layer between the screen and the application layer, as shown with the blue color on the right of Figure \ref{fig:overlay-structure}. The proxy layer first consumes and recognizes user input gestures. Then, it uses the gesture information to build ``virtual gestures'' (e.g., a swipe with a reversed direction). Finally, it passes the virtual gestures to the application layer.

To implement the proxy layer in Android, we use the Accessibility Service \cite{a2023_accessibilityservice}. We adopt a transparent Accessibility overlay as our proxy layer, and customize a Gesture Detector class to obtain the input gesture information, including the duration, location (or trajectory) and the number of fingers. To build a virtual gesture, we employ the \texttt{dispatchGesture} API. This API takes gesture information as parameter and performs the gesture on the current interface. Thus, we can easily manipulate gesture information to create the input interventions, as introduced in Section \ref{sub:system:design-space}.

We implement our input manipulation intervention techniques as follows:
\begin{itemize}
 \item Tap Delay: Add a delay \texttt{T$_{\textit{tap-delay}}$} to execute \\ \texttt{dispatchGesture} for taps;
 \item Tap Prolong: Set a threshold of tap duration \texttt{T$_{\textit{tap-threshold}}$} and ignore the taps whose duration is less than the threshold;
 \item Tap Shift: Add a pair of fixed shift values \texttt{(x,y)} to the coordinate of the users' single tap;
 \item Tap Double: Remap a double tap to a single tap, and ignore single taps. When the input is double tap, the output is single tap. Note that output cannot be double tap as the input double tap is remapped;
 \item Swipe Delay: Add a delay \texttt{T$_{\textit{swipe-delay}}$} to execute \\ \texttt{dispatchGesture} for swipes;
 \item Swipe Deceleration: Multiply the duration of \texttt{gesture} with a factor \texttt{F$_{\textit{swipe-decelerate}}$} so the effect of the swipe is decelerated;
 \item Swipe Reverse: Replace the trajectory of a swipe with a reversed one;
 \item Swipe Multiple Fingers: Set a threshold of a number of fingers \texttt{N} and ignore the swipes whose number of fingers is fewer than the threshold.
\end{itemize}

\paragraph*{Intervention Delivery Mechanism}
\label{sec:intervention-delivery-mechanism}

Our implementation allows users to set which apps they want to intervene and the usage limit. InteractOut monitors the usage time for each app via the Accessibility API.
When the limit of an application is reached, the proxy overlay will launch with selected interventions.

InteractOut is highly customizable. 
For interventions with thresholds, their intensity can be adjusted by changing the threshold. The configuration of the intervention and its intensity can also be customized for individual applications.
It is noteworthy that for the interventions with parameters \texttt{T$_{\textit{tap-threshold}}$}, \texttt{T$_{\textit{tap-delay}}$}, \texttt{T$_{\textit{swipe-delay}}$}, and \texttt{F$_{\textit{swipe-decelerate}}$}, we propose a dynamic design for the interventions to gradually take effect: the intervention intensity will start from zero and gradually increase to a target maximum value. Specifically, we apply a linear step increase on \texttt{T$_{\textit{tap-threshold}}$}, \texttt{T$_{\textit{tap-delay}}$} and \texttt{T$_{\textit{swipe-delay}}$}, and an exponential step increase on \texttt{F$_{\textit{swipe-decelerate}}$}. The step increase can be triggered by either a user operation (a tap, swipe, etc.) or the time elapsed since the last increase.
In this way, users will perceive very subtle interventions at first, which then become increasing stronger over time.

\paragraph*{Notification and Bypass Mechanism}
\begin{figure}[b!]
 \centering
 \includegraphics[width=.8\linewidth]{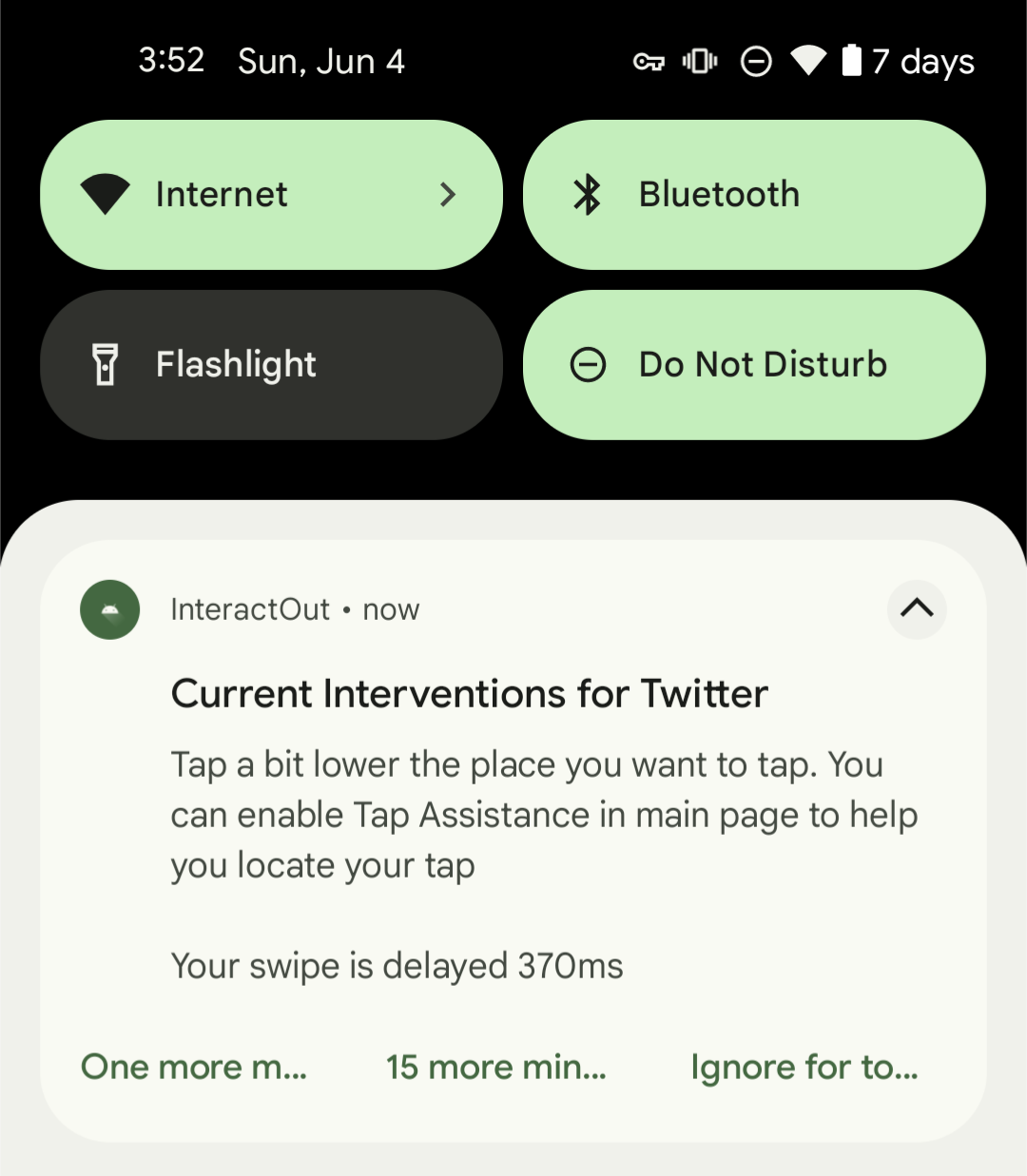}
 \caption{Bypass Option Menu in Android Notification Drawer. User can select 1 or 15 minutes or ignore the limit for the rest of the day directly in the notification.}
 \Description{Bypass option menu in Android notification drawer. It shows the current interventions and corresponding intensities. Users can choose from 1 more minutes, 15 more minutes and ignoring for today.}
 \label{fig:notification}
\end{figure}

InteractOut runs in the background and does not have an interface when delivered.
Instead, we leverage a persistent notification in the notification drawer \cite{notifications} to inform users of the intervention techniques and their intensity (if applicable). 
Explicit notifications are not necessary, as users will notice the input manipulations as they start to take effect.
Furthermore, following the implementation of the iOS Screen Time bypass options, we provide a bypass mechanism with the same options for users to select from three options (Figure \ref{fig:notification}) ``One more minute'', ``15 more minutes'', or ``Ignore for today'' directly through the notification drawer to pause the intervention. This notification is always available so that users can bypass the intervention at any time. We also offer these bypass options in the InteractOut app.

To finalize the InteractOut implementation, one key question remains: what should the intervention intensity be to effectively reduce smartphone overuse? 
We explore this in the next section.
\aptLtoX[graphic=no,type=html]{
\begin{table}[!t]
 \caption{The Two Intensity Levels Used in the Lab Study. Note that dp is a density-independent pixel whose actual length is based on the physical screen.}
 \label{tab:lab-study-config}
\begin{tabular}{c|c|cc}
\hline
\textbf{Gesture Category} & \textbf{Manipulation Strategy} & \textbf{Level 1} & \textbf{Level 2} \\ \hline
\multirow{4}{*}{Tap} & Delay & 500ms & 1000ms \\
 & Prolong & 100ms & 200ms \\
 & Shift & y+100dp & y+200dp \\
 & Double & yes & / \\ \hdashline
\multirow{4}{*}{Swipe} & Delay & 300ms & 800ms \\
 & Deceleration & $\times0.5$ & $\times0.25$ \\
 & Reverse & yes & / \\
 & Multiple fingers & 2 & 3 \\
\hline
\end{tabular}
\end{table}

\begin{figure}
 \centering
\caption{The Two Mobile Applications Used in the Lab Study. (a) Bubble Mania, for tap implementations. (b) Twitter, for swipe implementations.}
 \includegraphics[width=0.30\textwidth]{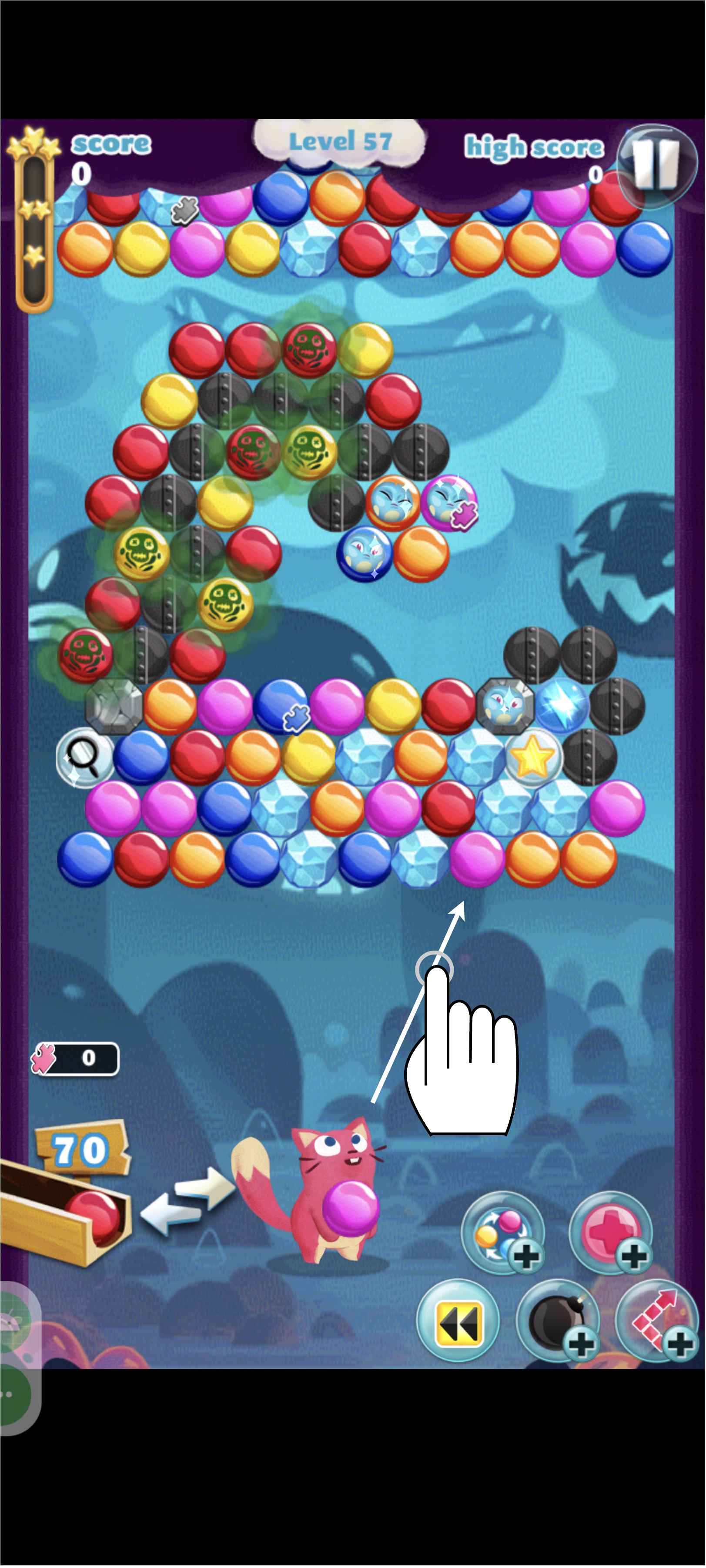}
 \Description{A gameplay screenshot of Bubble Mania, a typical bubble pop game where the user taps on the screen to shoot a bubble. Precise control is important in the game, while it is not time-sensitive.}
 \label{fig:bubble-mania-screenshot}
 \includegraphics[width=0.30\textwidth]{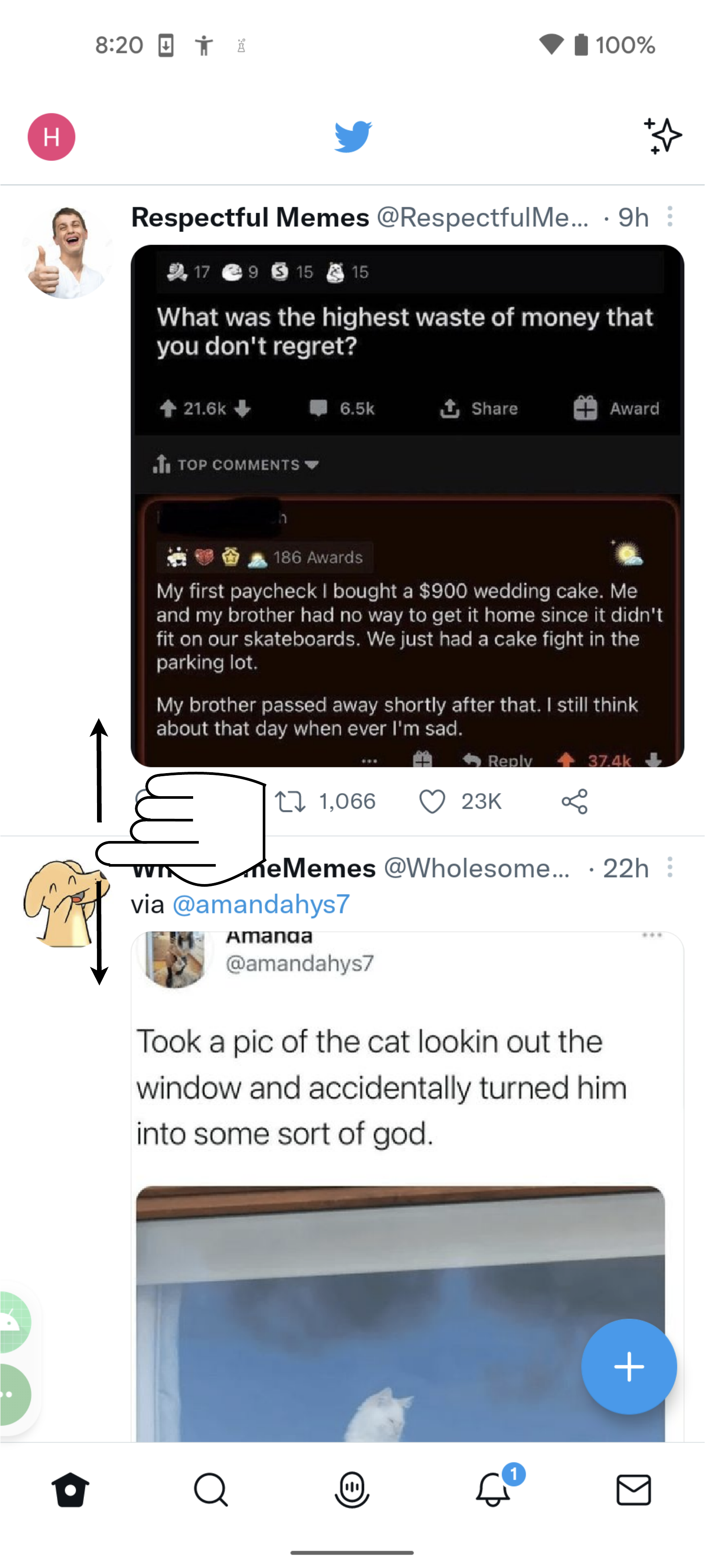}
 \Description{A screenshot of Twitter homepage. Scrolling is the main gesture when user browses tweets.}
 \label{fig:twitter-screenshot}
\end{figure}}{
\renewcommand{\arraystretch}{1.3}
\begin{table*}[!t]
\begin{minipage}[!t]{0.6\linewidth}
 \caption{The Two Intensity Levels Used in the Lab Study. Note that dp is a density-independent pixel whose actual length is based on the physical screen.}
 \label{tab:lab-study-config}
\resizebox{1\textwidth}{!}{
\begin{tabular}{c|c|cc}
\thickhline
\textbf{Gesture Category} & \textbf{Manipulation Strategy} & \textbf{Level 1} & \textbf{Level 2} \\ \hline
\multirow{4}{*}{Tap} & Delay & 500ms & 1000ms \\
 & Prolong & 100ms & 200ms \\
 & Shift & y+100dp & y+200dp \\
 & Double & yes & / \\ \hdashline
\multirow{4}{*}{Swipe} & Delay & 300ms & 800ms \\
 & Deceleration & $\times0.5$ & $\times0.25$ \\
 & Reverse & yes & / \\
 & Multiple fingers & 2 & 3 \\
\thickhline
\end{tabular}}
\end{minipage}\hfill
\begin{minipage}[!t]{0.37\linewidth}
 \centering
\captionof{figure}{The Two Mobile Applications Used in the Lab Study. (a) Bubble Mania, for tap implementations. (b) Twitter, for swipe implementations.}
 \includegraphics[width=0.48\textwidth]{figures/bubble_mania_screenshot.pdf}
 \Description{A gameplay screenshot of Bubble Mania, a typical bubble pop game where the user taps on the screen to shoot a bubble. Precise control is important in the game, while it is not time-sensitive.}
 \label{fig:bubble-mania-screenshot}
 \hfill
 \includegraphics[width=0.48 \textwidth]{figures/twitter_screenshot.pdf}
 \Description{A screenshot of Twitter homepage. Scrolling is the main gesture when user browses tweets.}
 \label{fig:twitter-screenshot}
\end{minipage}
\end{table*}
}
\renewcommand{\arraystretch}{1.0}

\section{Lab Study}
\label{sec:lab_study}
As introduced in Section~\ref{sec:system}, InteractOut can support various intervention implementations and intensities. This leads to the question of finding the right balance between restrictiveness (as over-restrictiveness leads to bad user experience) and flexibility (as over-flexibility leads to low engagement).
To answer this question, we first conducted a pilot lab study to compare different implementations of InteractOut and their intensity levels.
Our findings about the appropriate intervention and intensity selection will inform the field deployment experiment in Section~\ref{sec:field_study} to evaluate InteractOut in real-world settings.

\subsection{Participants}
\label{sec:lab_study_participant}

We recruited participants from our university's mailing lists. We asked participants to fill out a screening survey to evaluate their smartphone addiction level (SAS-SV \cite{SASSV}) and their desire to change smartphone usage behavior (TTM \cite{TMM}), and to collect their demographics information.
We set a SAS-SV score threshold of 30 (indicating smartphone-additive) and a TTM threshold of beyond stage (indicating the desire to change behavior).
From a pool of 112 survey respondents who met the inclusion criteria, we randomly sampled 30 participants (10 female, 20 male, aged 19 to 30).

\subsection{Design and Procedure}

We employed a within-subject design with a baseline condition of no intervention and our InteractOut condition with 8 gesture intervention implementations (see Section \ref{sub:system:design-space}). For each implementation, each of them was assigned two different intensity levels (see Table \ref{tab:lab-study-config}). These levels were chosen based on UI design principles (e.g., important limits of response times~\cite{nngroup}) and early pilot studies.
We chose Twitter (now known as X) \cite{Twitter} and Bubble Mania \cite{BubbleMania}, two popular and addictive apps \cite{AddictiveMobileApps}, which represented apps with swipe-heavy and tap-heavy interactions, to mimic realistic smartphone usage scenarios.
We used a Google Pixel 5a running Android 12.

At the beginning of our study, participants were asked to complete a pre-study survey to collect commerical intervention products they have used.
They completed an interactive tutorial to familiarize themselves with the 8 gesture intervention implementations.
Then, participants were asked to test each technique for 3 minutes on the two assigned apps. The order of implementations was counterbalanced to minimize order effects.
After each implementation stage, participants were asked to complete a short survey on a 7-point Likert scale regarding their perception of each implementation. The questions include annoyance level, disruptiveness, usage desire reduction, mental load, and physical load. Moreover, if applicable, we asked participants which intensity level they would prefer to use in daily life.
After going through all implementations in one category (tap or swipe), they were asked to rank the four manipulation strategies based on perceived effectiveness and receptivity in reducing the desire to continue using the app.
Finally, we conducted an exit interview about their experience using these implementations and general comments. Our study was approved by our institution's IRB and lasted about 2 hours. Participants were compensated with \$30 for their time.

\subsection{Results}

\begin{figure*}[t!]
\centering
\begin{subfigure}{0.49\textwidth}
 \centering
 \includegraphics[width=1\linewidth]{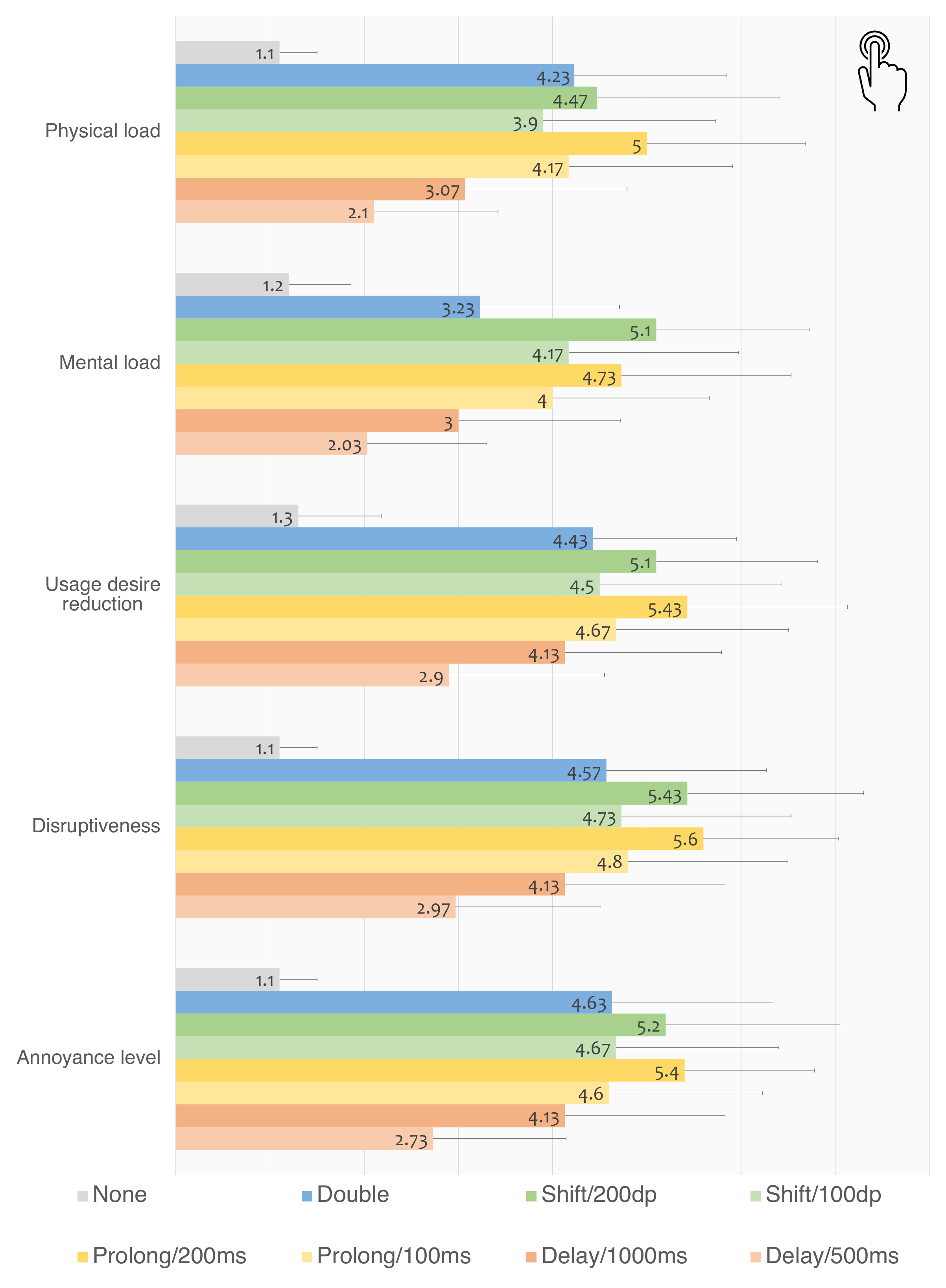}
 \caption{}
\end{subfigure}%
\hfill
\begin{subfigure}{0.49\textwidth}
 \centering
 \includegraphics[width=1\linewidth]{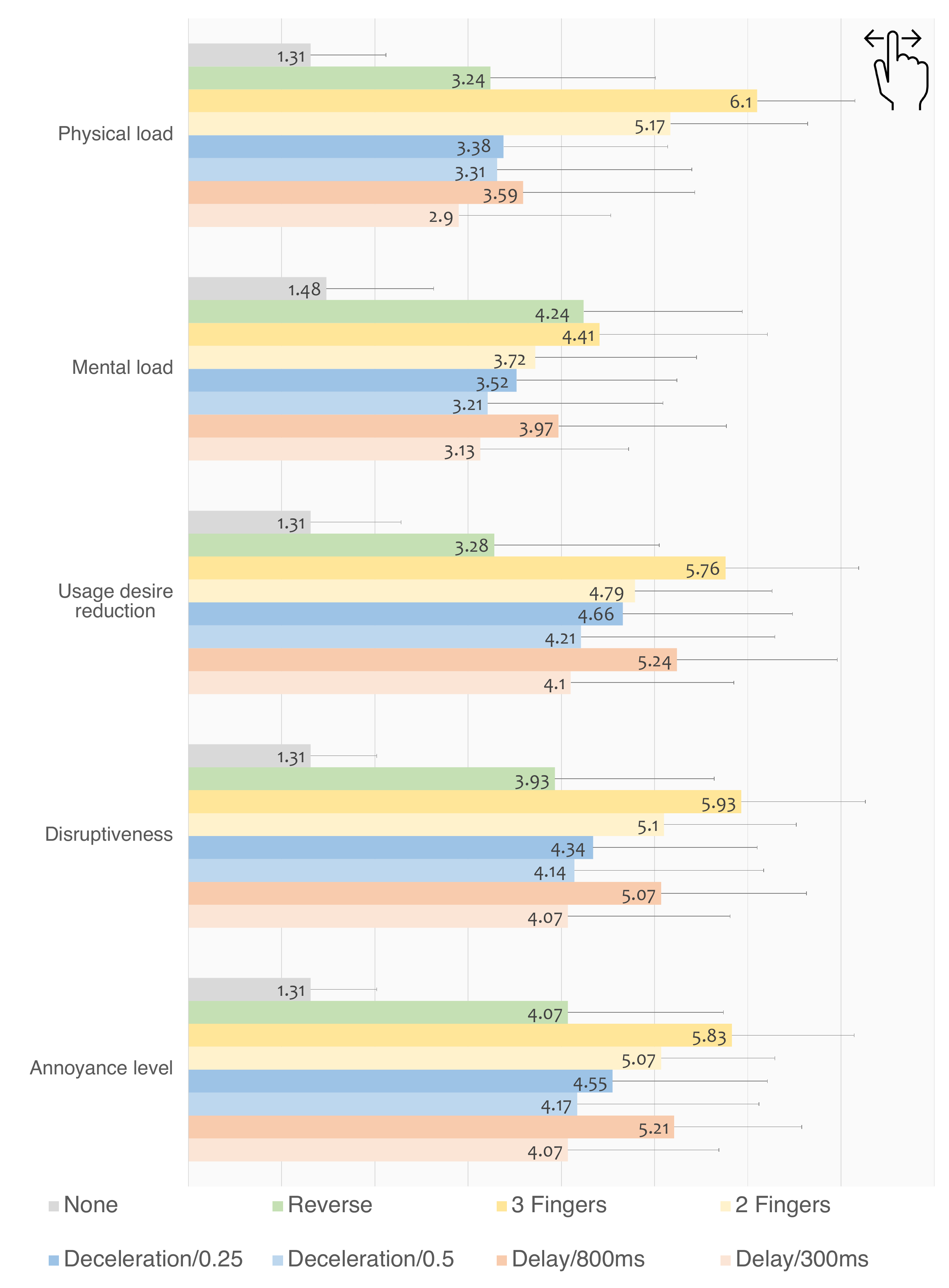}
 \caption{}
\end{subfigure}
\vspace{-.5pc}
\caption{Participants' Likert Scale Responses on (a) Tap Implementations and (b) Swipe Implementations on a 7-point Likert Scale (1 - Strongly disagree and 7 - Strongly agree). As expected, InteractOut are significantly more annoying, more disruptive, more mentally and physically demanding than the baseline condition with no intervention.}
\label{fig:intervention-mean-sd}
\Description{Clustered bar chart showing mean and standard deviation of participants' Likert scale responses on tap, swipe, and no interventions. The result is shown for the first five questions and implementations with different intensity levels. For tap implementations, the results for each question are close, where no intervention has significantly lower score than other implementations. Tap Delay is also less annoying or disruptive than Tap Prolong, Tap Shift, and Tap Double. For swipe implementations, each question has similar results. Swipe Delay, Swipe Scale, and Swipe Reverse have approximately the same average while Swipe Multiple Fingers scores higher and no intervention has a much lower score.}
\end{figure*}

\begin{figure}
\centering
\begin{subfigure}{\linewidth}
 \centering
 \includegraphics[width=\linewidth]{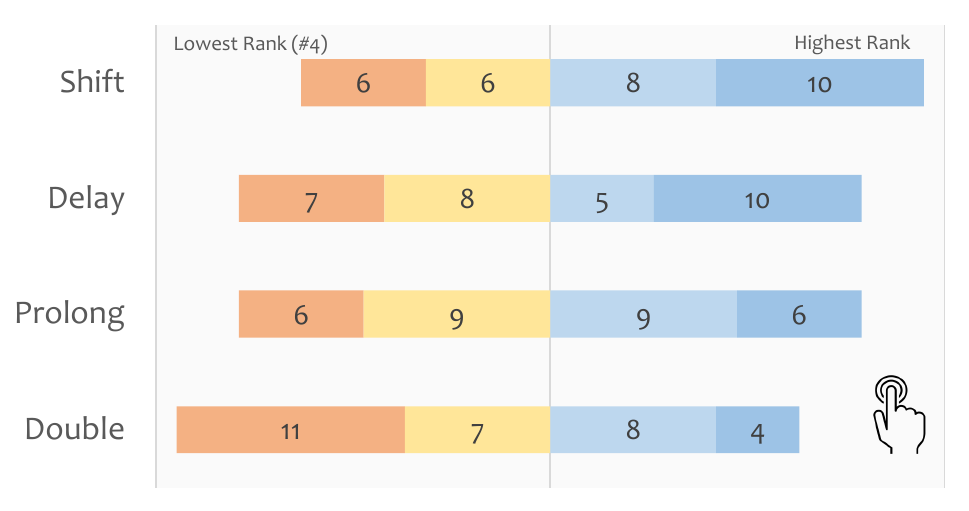}
 \caption{}
\end{subfigure}%
\hfill
\begin{subfigure}{\linewidth}
 \centering
 \includegraphics[width=\linewidth]{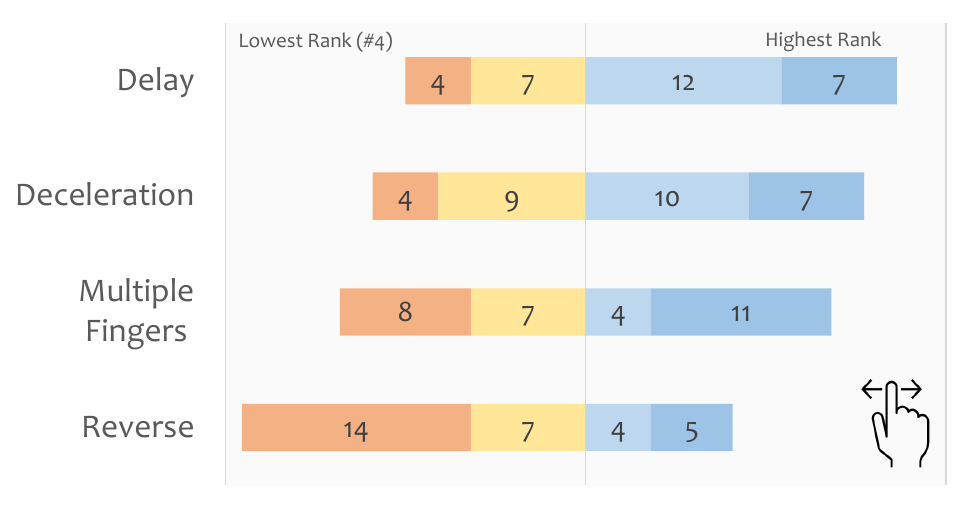}
 \caption{}
\end{subfigure}
\vspace{-1.8pc}
\caption{Participants' Perceived Effectiveness and Receptivity of Manipulation Strategies within (a) Tap Implementations and (b) Swipe Implementations. Tap Shift and Swipe Delay were the most preferred, while Tap Double and Swipe Reverse were the least preferred.}
\label{fig:rank}
\Description{Diverging stacked bar chart showing participants' rankings on tap and swipe implementations. For tap implementations, Tap Shift has the top ranking with 10 participants rank first and 8 rank second. It is followed by Tap Delay and Tap Prolong, and finally Tap Double with only 4 participants rank top. For Swipe implementations, Swipe Delay is more favored than other implementations, with 7 participants rank first and 12 rank second. Swipe Scale is slightly behind with 2 less participants rank it second. They are then followed by Swipe Multiple Fingers. And Swipe Reverse is least favored with only 5 participants rank top and 4 rank second.}
\vspace{-1pc}
\end{figure}

Figure \ref{fig:intervention-mean-sd} presents the questionnaire results, and Figure \ref{fig:rank} summarizes the rankings results.
Not surprisingly, we found that both tap and swipe implementations were significantly more effective than the baseline with no interventions (both $p < 0.0001$).
We also found that participants generally preferred the higher intensity level of Tap Delay, Tap Shift, Swipe Delay, and Swipe Deceleration.

\subsubsection{Implementation Intensity Levels Comparison}
\begin{figure}
\centering
\begin{subfigure}{\linewidth}
 \centering
 \includegraphics[width=\linewidth]{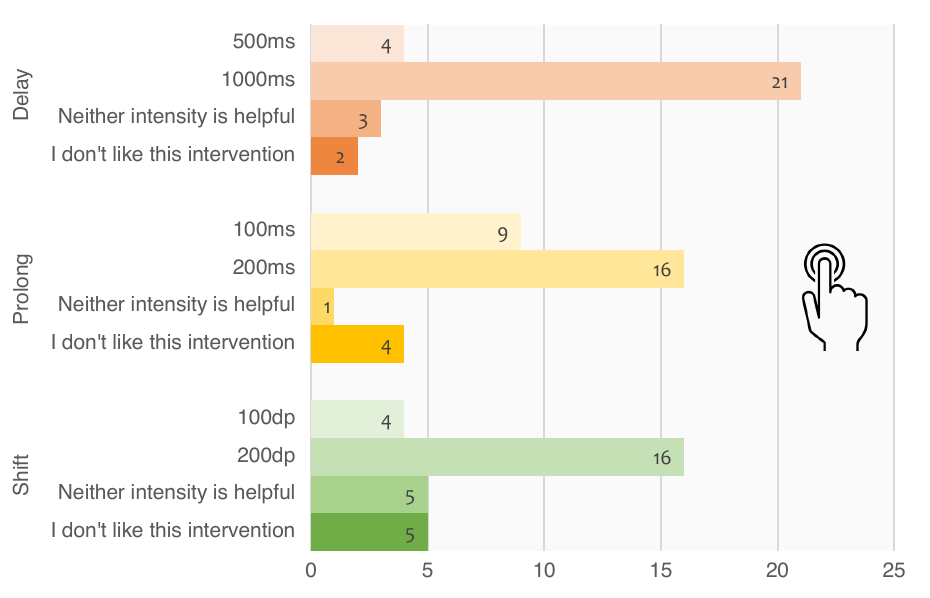}
 \caption{}
\end{subfigure}%

\begin{subfigure}{\linewidth}
 \centering
 \includegraphics[width=\linewidth]{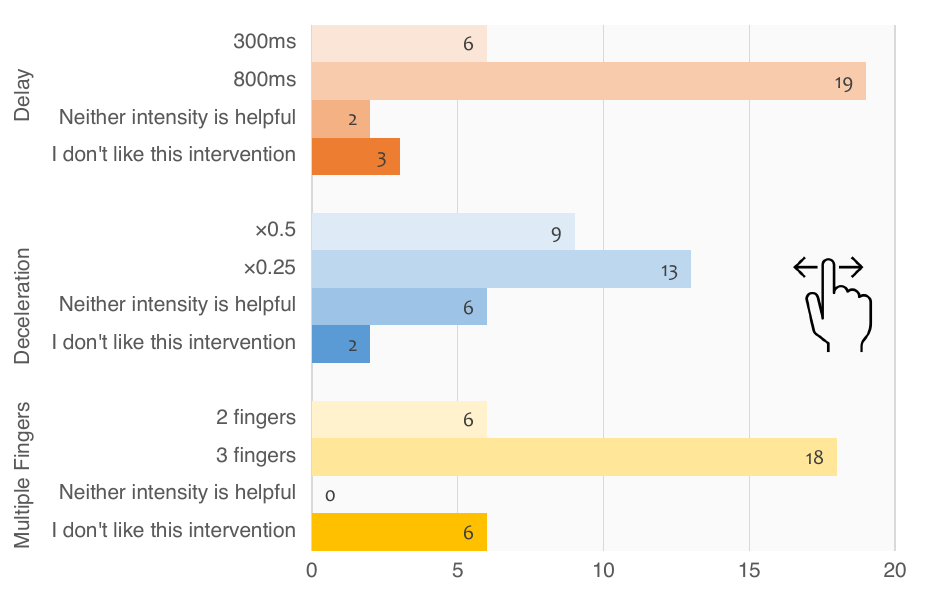}
 \caption{}
\end{subfigure}
\vspace{-1.9pc}
\caption{Participants' Preferences on the Two Intensity Levels of (a) Tap Implementations and (b) Swipe Implementations. Higher intensities were always the preferred choice for all implementations.}
\label{fig:level-pref}
\Description{Clustered bar chart showing participants' preferences on the two intensity levels of tap and swipe implementations. The participants favored higher intensities for all implementations. Out of 30 participants, there were respectively 21, 16, and 16 participants choosing higher intensities of Tap Delay, Tap Prolong, and Tap Shift. Meanwhile, 19, 13, and 18 participants preferred higher intensity levels for Swipe Delay, Swipe Scale, and Swipe Multiple Fingers.}
\end{figure}

\textbf{We found that most participants preferred the higher intensity level for most implementations} (Figure \ref{fig:level-pref}), even though the high-intensity level could be more disruptive and demanding. Participants could easily ignore the implementations with the lower intensity level: \textit{``500ms delay is too short that I could not even tell if there is intervention or not''} (P21), \textit{``I need to put extra effort to (successfully) tap the screen when it's 200ms prolonged compared to 100ms''} (P4).
However, one exception was for Swipe Deceleration. Between the $\times0.5$ and $\times0.25$ levels, participants could hardly tell the difference, as shown in the Likert scale responses ($p = 0.42$) and preferences (9 vs. 13).
In summary, as shown in Figure~\ref{fig:intervention-mean-sd} and Figure~\ref{fig:level-pref}, the higher intensity levels we chose tend to be more effective and fairly acceptable.
Thus, in the next subsection, we only refer to the higher intensity level if not explicitly stated.

\subsubsection{Intervention Implementations Comparison}

We found that the effectiveness and receptivity of an implementations is related to its physical and mental loads. \textbf{Implementations with a higher mental load but a lower physical load were the most preferred.}
They were effective because participants needed to think before taking actions, but they were not too annoying because it took no more physical effort than normal actions once participants got used to it.
For example, P4 and P12 both mentioned that the Tap Shift in Bubble Mania and other similar use cases enlarges the gulf of execution by requiring additional cognitive load to think about where to activate.

Meanwhile, the less preferred implementations were those with high physical load.
For instance, the Swipe Multiple Fingers was extremely physically demanding. Many participants had to adopt unnatural postures to swipe with multiple fingers.
Interestingly, despite its strong annoyance, some participants still ranked it as the most preferred implementation because it was effective yet quite intuitive (thus a low mental load).
The least effective implementation was Swipe Reverse, which has both low physical and mental loads. Most participants made several mistakes at first, but they could easily get used to it within a few trials: \textit{``It reminds me of my first use of the MacBook trackpad''} (P22). 

Through the lab study, we identified that among the methods we tested, the higher intensity level of Tap Delay, Tap Shift, Swipe Delay, and Swipe Deceleration are the four implementations that are better at striking the balance between effectiveness and user experience.
However, due to the short experiment duration during the lab study, it is unclear whether these four implementations will still be as effective in real-world settings. Furthermore, how users balance mental and physial load could be different in lab vs. field settings.
Therefore, the lab study not only informed the implementation and intensity selection and design of our intervention app, but also motivated us to deploy our app in the real world.

\section{Field Experiment}
\label{sec:field_study}
Based on the lab study, we developed an intervention app consolidating the four preferred intervention techniques.
We then conducted a five-week field experiment. The goal of the field experiment is to understand how InteractOut influences smartphone usage and user receptivity in real-world scenarios.
We also compare InteractOut against one of the most common intervention methods, the Timed Lockout intervention.
Our results indicate that InteractOut outperforms the Timed Lockout with significantly more smartphone usage reduction and a higher intervention acceptance rate.

\subsection{Intervention Techniques}
We compare InteractOut against a baseline intervention technique Timed Lockout, a popular technique widely available on modern smartphones.

\subsubsection{InteractOut Interventions.}
Our final version of the intervention includes the four implementations with the high-intensity level: Tap Delay (1000 ms), Tap Shift (y-200 dp), Swipe Delay (800 ms), and Swipe Deceleration ($\times$0.25). All interventions except for Tap Shift follows the step increase feature we discussed in Section \ref{sub:mobile}. Every time users touch the screen, the intervention intensity increases by a small step (10 ms for Tap/Swipe Delay and $4^{-\frac{1}{100}}$ for Swipe Deceleration). This means that the intervention will saturate within 100 touches for Tap Delay and Swipe Deceleration and 80 touches for Swipe Delay. If users do not touch the screen for one minute, the intensity will also increase by one step.

Our app provides a config stage, where users can choose the apps that they want to receive intervention as well as the time budget (see Figure~\ref{fig:app_choose} and \ref{fig:time_setting}).
In our field experiment, we fix the daily time budget as one hour in total for all target apps combined to ensure consistent control in our study, and for most participants to reach the limit and experience the intervention. But in practice, users can adjust the budget based on their needs.
The bypass menu of InteractOut is shown in Figure \ref{fig:notification}. It is also available on the main page of the InteractOut app.

\subsubsection{Baseline: Timed Lockout.}
\begin{figure*}[t!]
\centering
\begin{subfigure}{.24\textwidth}
 \centering
 \includegraphics[width=0.7\linewidth]{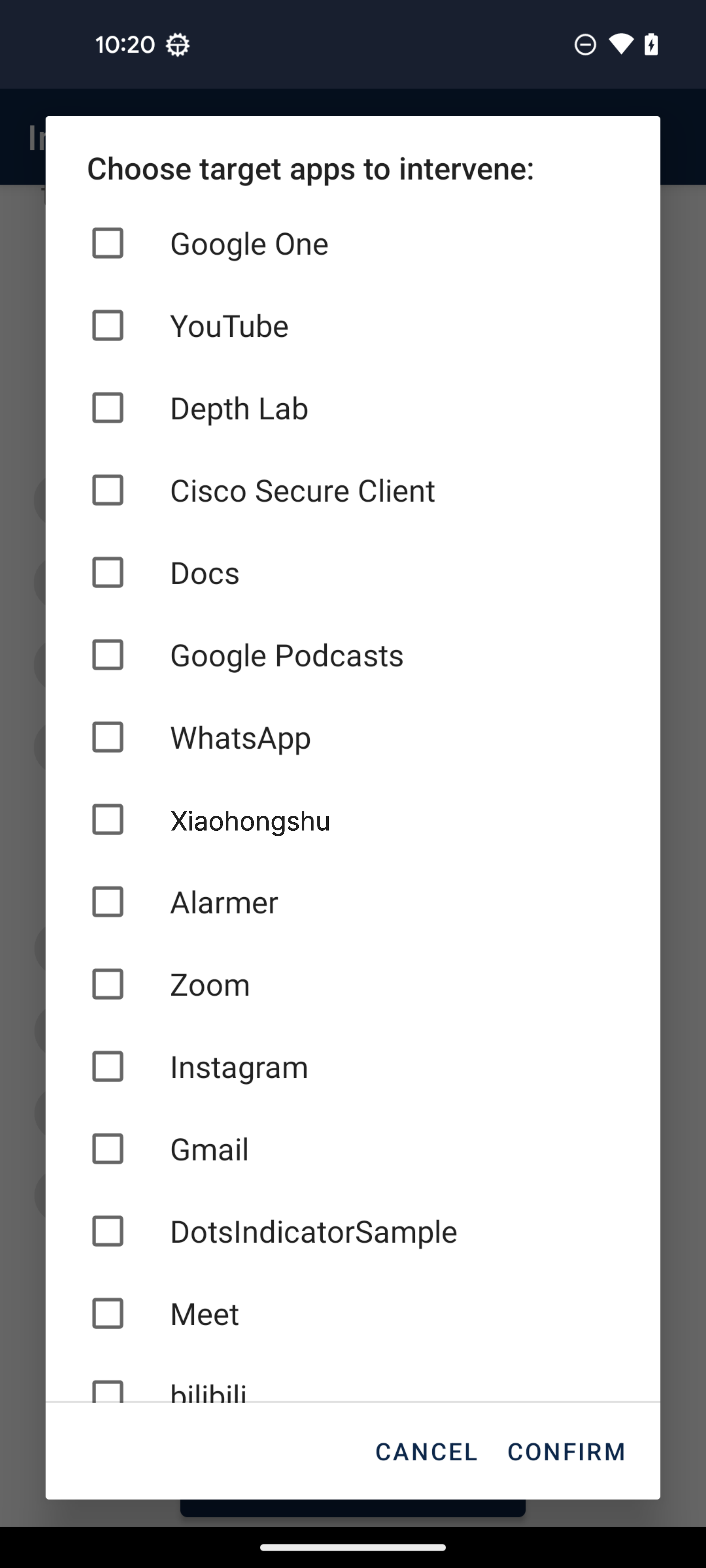}
 \caption{}
 \Description{Target app selection page of InteractOut, user can choose the app they want to restrict the time.}
 \label{fig:app_choose}
\end{subfigure}%
\begin{subfigure}{.24\textwidth}
 \centering
 \includegraphics[width=0.7\linewidth]{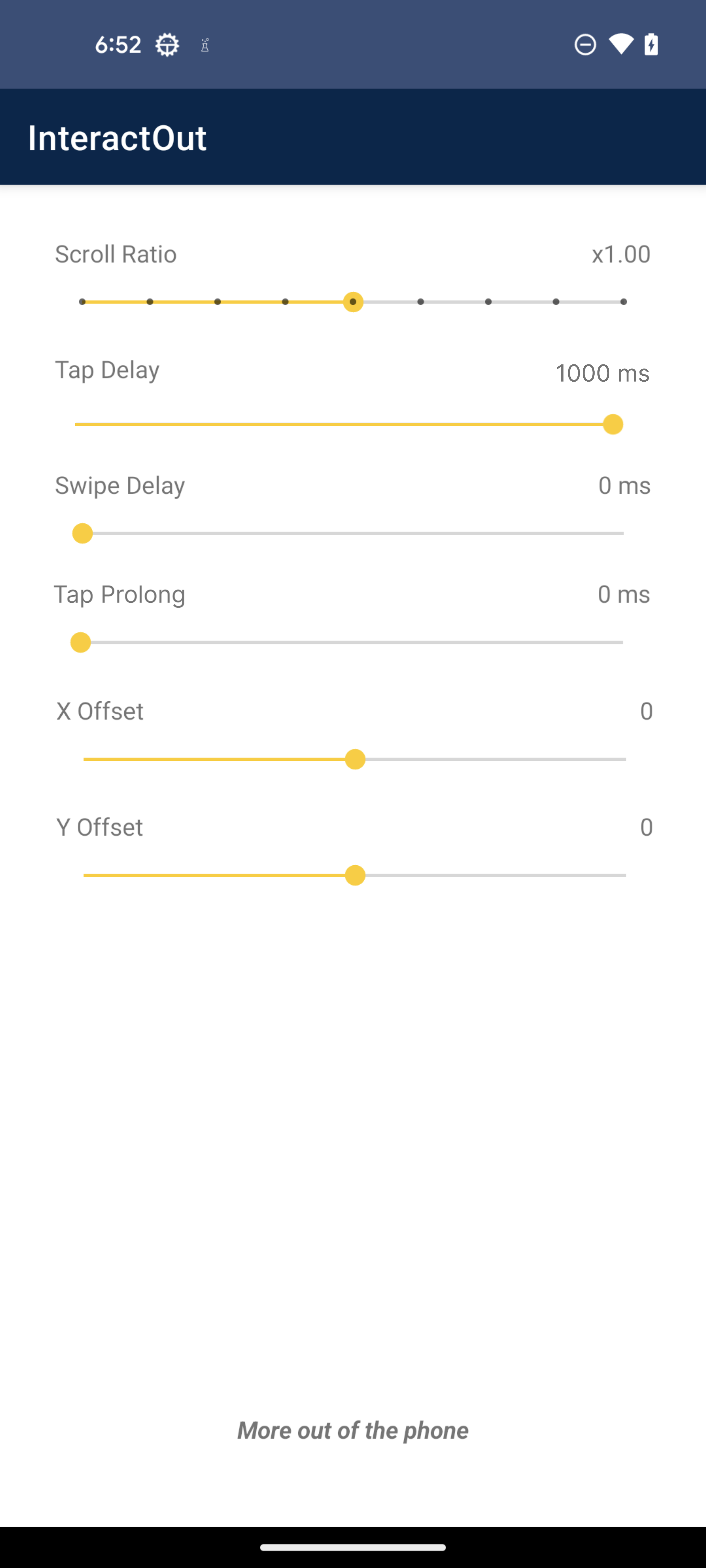}
 \caption{}
 \Description{Configuration page of InteractOut, user can drag the sliders to adjust the intensity of different interventions.}
 \label{fig:config_page}
\end{subfigure}%
\begin{subfigure}{.24\textwidth}
 \centering
 \includegraphics[width=0.7\linewidth]{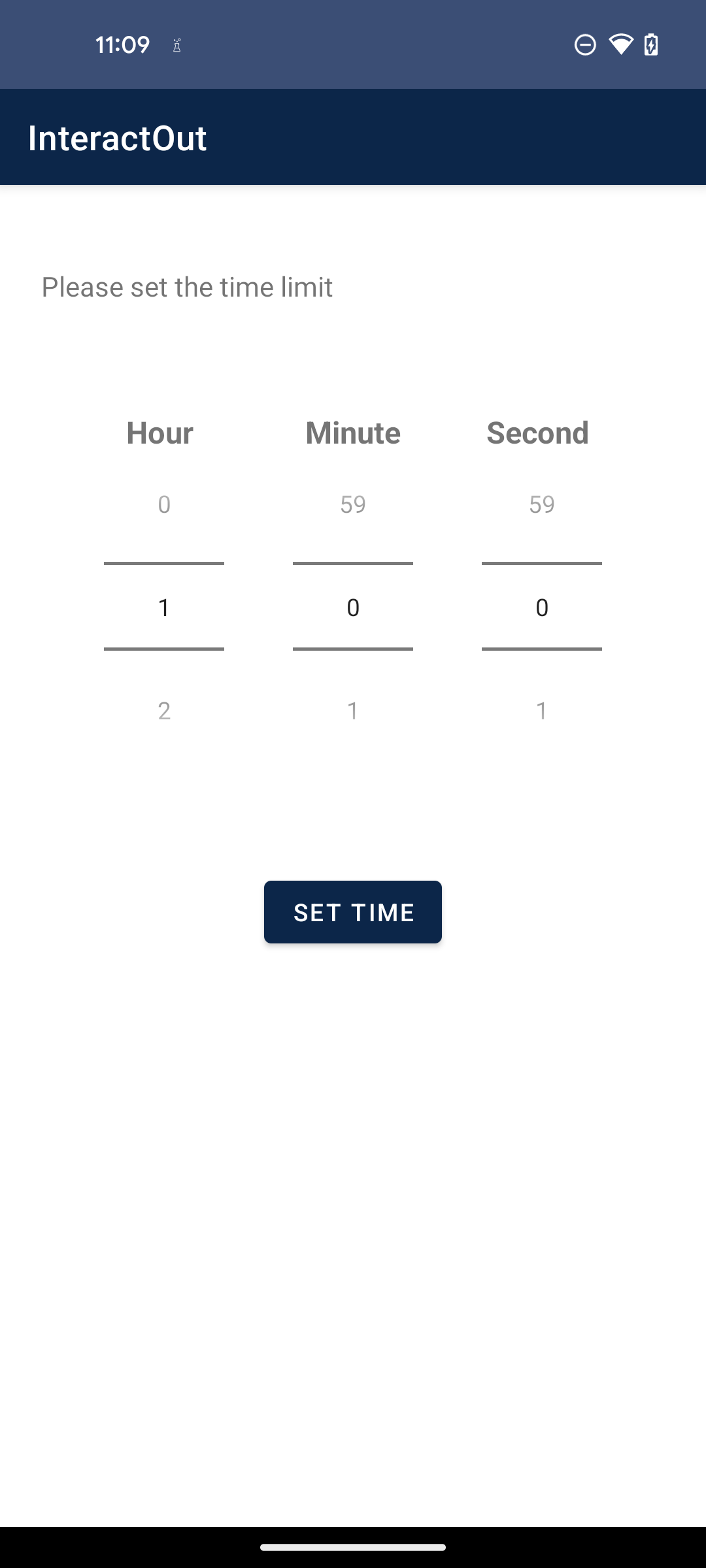}
 \caption{}
 \Description{Time setting page of each target app. }
 \label{fig:time_setting}
\end{subfigure}%
\begin{subfigure}{.24\textwidth}
 \centering
 \includegraphics[width=0.7\linewidth]{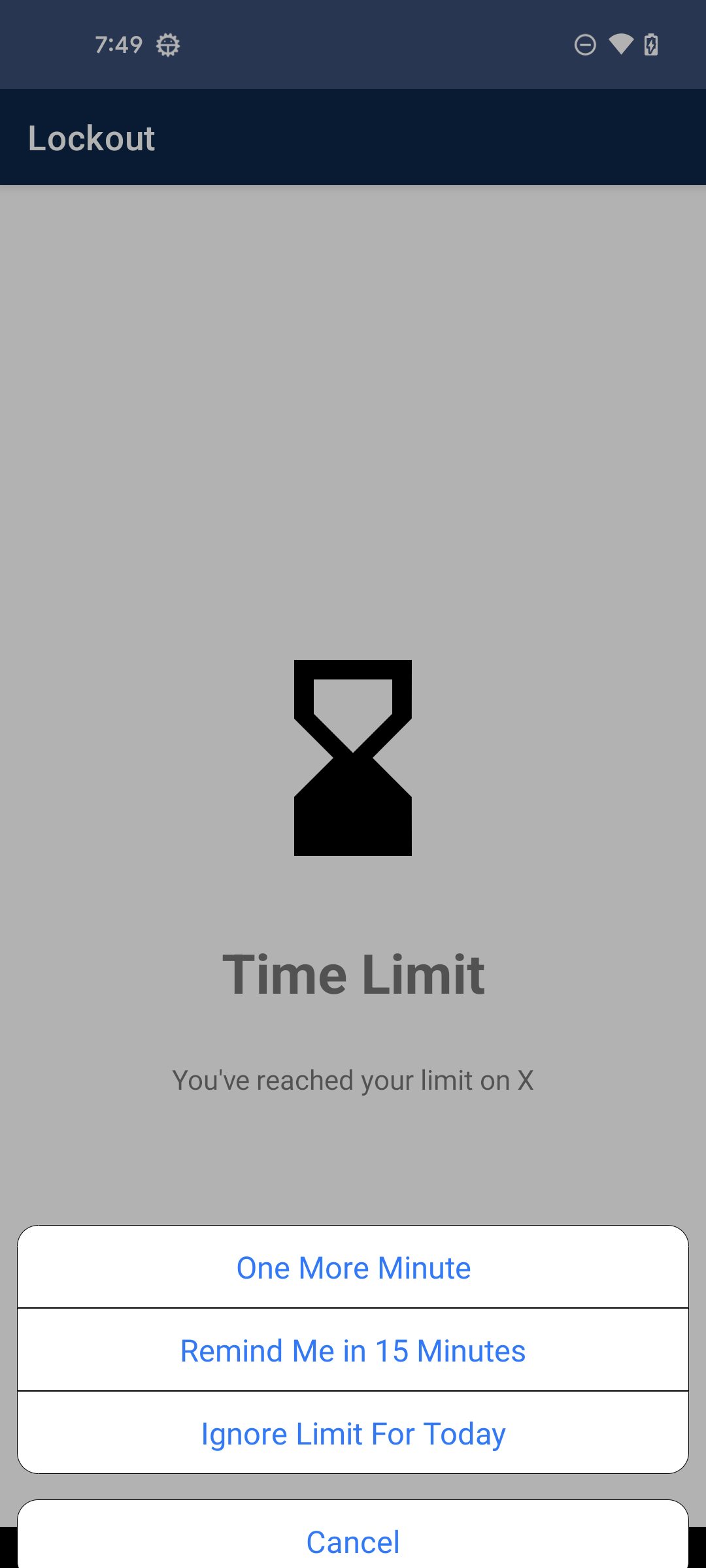}
 \caption{}
 \Description{Bypass menu of the lock screen. There are 3 options: "One More Minute", "Remind Me in 15 Minutes", and "Ignore Limit For Today", together with a "Cancel" button to close the menu.}
 \label{fig:field_lockout_bypass}
\end{subfigure}
\vspace{-.8pc}
\caption{(a) Target App Selection Page. Users specify the target apps to intervene during the onboarding stage of the field experiment; (b) InteractOut's Configuration Page for Interventions and Intensities {and (c) InteractOut's Configuration Page for Total Time Limit} -- Note that users cannot change these configurations for the purpose of the study. We discuss future use cases in Section~\ref{subsec:discussion_future_deployment}; and (d) Lockout's Bypass Menu, triggered by selecting ``Ignore Limit'' in the popup blocking screen. {Note that InteractOut's bypass menu is shown in Fig.~\ref{fig:notification}.}}
\end{figure*}

Timed Lockout is one of the most common intervention techniques supported by modern Android and iOS operating systems~\cite{android_screentimer,ios_screentimer}. 
Users can set time limits based on individual apps or groups of apps. Once the limit is reached, the Timed Lockout intervention will pop out a blocking interface notifying the user as well as providing options to bypass the intervention to request more usage (see Figure~\ref{fig:field_lockout_bypass}).
For our study, we also fixed the time limit to be one hour total, making it consistent with InteractOut.
For simplicity, we use ``Lockout'' as an abbreviation of the Timed Lockout intervention.

\subsection{Participants}
We recruited a new group of participants from our university's mailing lists. Similar to the lab study, we used the same screening survey and inclusion criteria. Moreover, we required participants to be regular Android users.
From a pool of 69 survey respondents who met the inclusion criteria, a total of 42 participants joined and completed the field experiment (11 female, 31 male, aged 18 to 34). Their Android device manufacturers include Samsung (23), Google (11), OnePlus (2), Xiaomi (2), Motorola (3) and LG (1). None of them participated in the previous lab study.

\begin{figure}[b!]
 \centering
 \includegraphics[width=\linewidth]{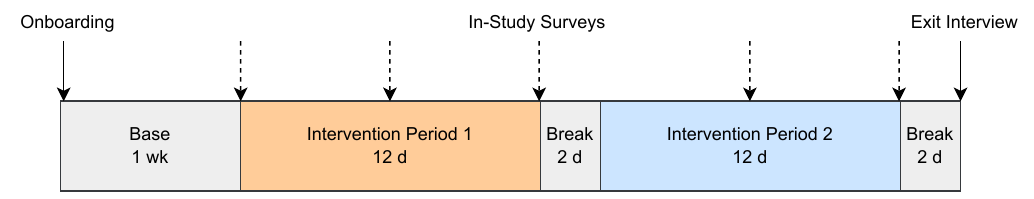}
 \caption{Field Experiment Schedule. The order of the two interventions is counterbalanced. The break days are inserted to investigate the lasting effect of interventions and to reduce the influence between interventions. {There are 5 in-study surveys in total, 1 at the end of the Base week and 2 for each intervention period, with one in the middle and one at the end.}}
 \Description{The field experiment is 5 weeks long. The first stage is a base week with no intervention. The second stage is intervention period 1, with 12 intervention days and 2 break days with no intervention. The third stage is the same as the second stage, except that the intervention is different. The participant used the Lockout intervention or InteractOut gesture intervention in the first intervention period and used another in the second intervention period.}
 \label{fig:field_study_schedule}
\end{figure}

\subsection{Design}
\label{sub:field_experiment:design}

\subsubsection{Experimental Design}
During the experiment, we adopted a within-subject design.
The 5-week schedule of the field experiment is shown in Figure \ref{fig:field_study_schedule}.
The first week was used for the base measurement {of participants' usual smartphone usage data} without any intervention, followed by 2 two-week intervention stages, each including 12 intervention days and 2 break days. {The break days were inserted to investigate the lasting effect of interventions (see Sec \ref{subsubsec:lasting_effect}) and to reduce the influence between interventions.} {Participants used one intervention in the first intervention stage and used the other in the second stage.} The order of the intervention technique was counterbalanced.
For InteractOut, since combining the two tap interventions or the two swipe interventions may introduce overly high interaction friction, we randomly picked one tap and one swipe intervention for each participant, resulting in four intervention combinations.

\subsubsection{Evaluation Metrics}
Our evaluation metrics focus on the app usage (usage time and opening frequency) change and acceptance rate of the intervention methods.
Thus, we logged app enter and exit events, as well as the instances of intervention encounter and bypass on participants' devices throughout the study.
The data collection was done through the Google Firebase database system \cite{google_2023_firebase}.
Each week, we also delivered an in-study survey to collect workload using NASA TLX \cite{thenasatlxtooltaskloadindex_2019_tlx}, and subjective feedback on the two interventions. At the end of the study, we conducted an exit interview for each participants for their overall perception of 2 intervention techniques.
At the end of the study, participants attended an offboarding session during which we conducted a semi-structured exit interview, and uninstalled the study app from the participants' devices.

\begin{figure*}[t!]
 \centering
 \includegraphics[width=\linewidth]{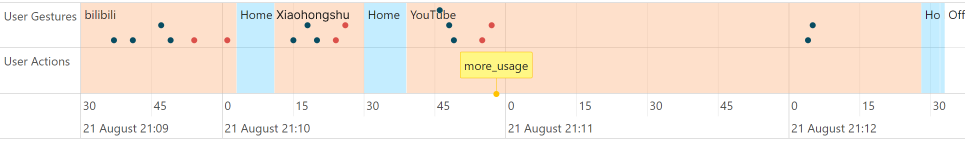}
 \Description{A sample app activity timeline under InteractOut. It shows that the user first used bilibili, then entered home screen, then entered Xiaohongshu, then went back to home screen, then entered Youtube and finally go back to home screen. During using bilibili, Xiaohongshu and YouTube, the user had some taps and swipes when InteractOut launched.}
 \label{fig:interactout_timeline}
\vspace{-1pc}
\caption{Web-Based Visualization Tool for App Activity Visualization. This example shows a sample app activity timeline under InteractOut. Black and red dots indicate swipes and taps during the intervention.
This tool serves two purposes: (1) to provide data collection transparency to end users and help them better understand the data, and (2) to help researchers better monitor study compliance and troubleshoot potential issues. Gesture events are only logged when intervention is on.}
\label{fig:vis}
\end{figure*}

\subsubsection{Visualization Tool}
Moreover, to better understand the data and control study compliance, we developed a web-based visualization tool (Figure \ref{fig:vis}) to show real-time app activities and daily activity summaries.
We used this tool to keep track of participants' app activities and troubleshoot outlier patterns (e.g., a multi-day gap between two app events).
During onboarding, we also used this tool to show participants their live data we were collecting, which provided them with data collection transparency. 

\subsection{Field Experiment Procedure}
Our study was conducted on a rolling basis from March 8th to April 28th, 2023.
In the onboarding session, participants {came to our lab,} installed our app, learned the features of their assigned gesture interventions, got familiarized with the visualization tool, and specified the target apps to apply screentime interventions. {The most frequently selected target apps included Instagram (25/42 participants), YouTube (18/42), Reddit (10/42), Discord (8/42) and Twitter/X (7/42). To manage the study, researchers used a Python script to automatically monitor usage activities and switch study phases, while manually sending out surveys according to a fixed schedule (Figure~\ref{fig:field_study_schedule}).}
After five weeks of the intervention period, participants went through the exit interview and ended the study.
Our study was approved by our institution's IRB.
Participants were compensated with up to \$100 USD based on the number of in-study surveys they completed and the days data were uploaded.
\section{Field Experiment Results}

During the five-week study, we collected a total of 852,556 app activities, 15,042 intervention encounters,
2,671 intervention bypasses, and 214,125 interactions during InteractOut intervention active period (single tap, double tap, and swipe). We aggregated these events to analyze each participant's app usage behavior and intervention acceptance rate. We analyzed the quantitative data and the qualitative data collected via questionnaires and interviews.

\subsection{App Usage Behavior}
\label{sec:app_behavior}
The primary goal of our intervention is to reduce app usage. Thus, we first investigate the app usage behavior: app usage time and app opening frequency.
Due to the large app usage variation among individuals, we normalized each participant’s data by calculating the ratio against their own data during the base week when there was no intervention. A ratio smaller or greater than 1 indicated that participants reduced or increased app usage compared to their ordinary behavior.
Our results showed that participants had a significantly smaller ratio when using InteractOut compared to the Lockout baseline on their target app usage behavior. 

\subsubsection{App Usage Time}
\label{sec:app_usage_time}
\begin{figure*}
\centering
\begin{subfigure}{.33\linewidth}
\centering
\includegraphics[width=\linewidth]{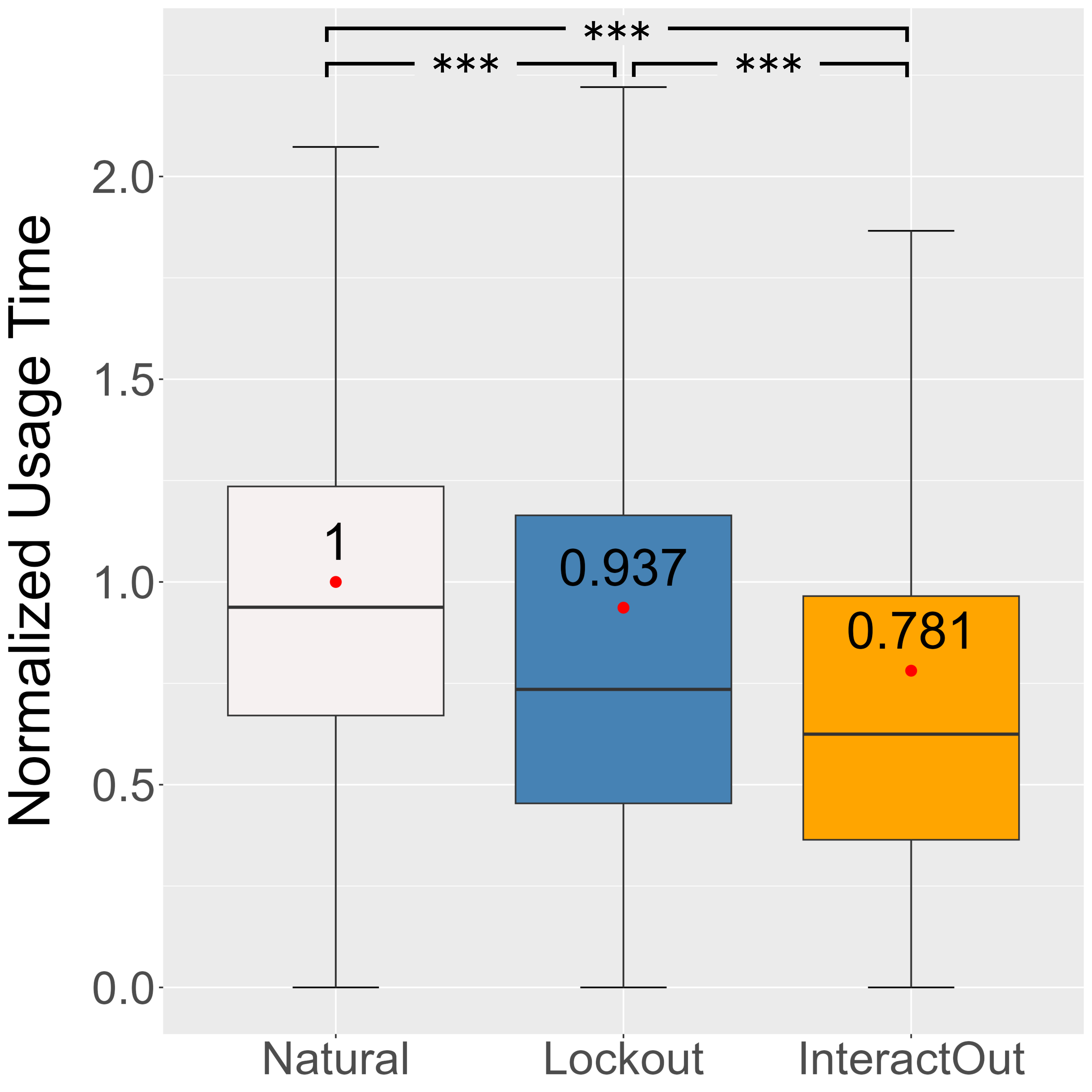}
\caption{Target Apps}
\Description{Boxplot of normalized target app usage time. The mean value of Natural (no intervention), Lockout and InteractOut is 1, 0.937 and 0.781. All of 3 pairwise comparisons show significance.}
\label{fig:target_usage}
\end{subfigure}%
\hspace{5pc}
\begin{subfigure}{.33\linewidth}
\centering
\includegraphics[width=\linewidth]{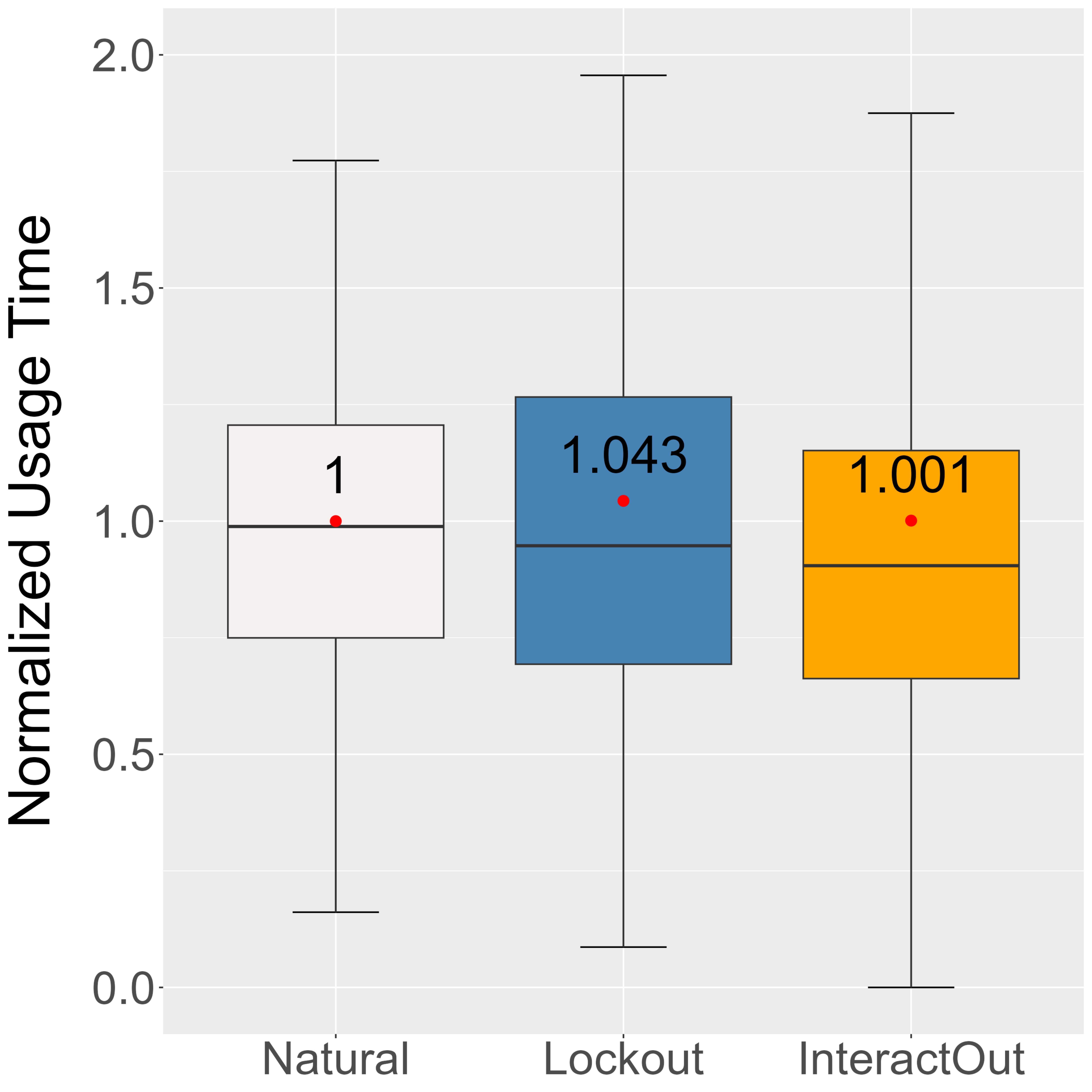}
\caption{Overall Usage}
\Description{Boxplot of normalized overall usage time. The mean value of Natural (no intervention), Lockout and InteractOut is 1, 1.043 and 1.001. None of 3 pairwise comparison shows significance.}
\label{fig:total_usage}
\end{subfigure}
\caption{Normalized App Usage Time in the Intervention Conditions. It shows the normalized ratio of the usage time against the average usage time in the base week. Lower ratio means more effective reduction. In each box, the middle black line shows the median and the red point shows the mean. The same apply to other box plot figures. Note that the minimum usage time is 0 as some participants do not use their target apps in some days.}
\label{fig:usage_time}
\end{figure*}

\textbf{Both InteractOut and Lockout significantly decreased participants' target app usage compared to naturalistic usage, while InteractOut had a significantly larger time reduction.} The average usage time was $149.04 \pm 7.24$ min in the base week and $131.20 \pm 5.56$ min, $103.70 \pm 4.67$ min for Lockout and InteractOut, respectively.
We then used the normalized data for comparison. As Figure \ref{fig:usage_time} shows, participants had a large usage time reduction of target apps during the InteractOut period (Mean = 0.78, Std = 0.029), which is 15.6\% lower than Lockout (Mean = 0.94, Std = 0.031).
A Shapiro-Wilk normality test showed that the usage time did not follow a normal distribution.
Therefore, we used a General Linear Mixed Model (GLMM) with a Log-Gamma link function (based on a Kolmogorov-Smirnov test) and participant ID as a random effect. The fixed effects were \textit{Intervention} (with levels ``Natural'', ``InteractOut'' and ``Lockout''), \textit{Order} (with levels ``InteractOut first'' and ``Lockout first''), their interaction (\textit{Intervention} $\times$ \textit{Order}), and \textit{Implementation} (with four levels being the 2$\times$2 combinations of tap/swipe interventions, as noted in Section \ref{sub:field_experiment:design}). Results showed that \textit{Intervention} was the dominant factor of the usage time change ($\chi^2 = 66.2$, $p_{Intervention} < 0.001$), while all other factors did not show significance ($p_{Order} = 0.52$, $p_{Intervention \times Order} = 0.21$, $p_{Implementaion} = 0.72$).
It is worth noting that the four combinations of input intervention techniques achieved similar results, indicating their robustness.
A follow-up pairwise post-hoc Tukey's HSD test on \textit{Intervention} further showed that both InteractOut ($Z = -8.52$, $p < 0.001$) and Lockout ($Z = -4.03$, $p < 0.001$) lowered the usage time, while InteractOut had a significantly lower usage time than Lockout ($Z = -4.41$, $p < 0.001$).
However, when looking at the usage time of all apps, we observed no significance ($p_{Intervention} = 0.13$, $p_{Order} = 0.86$, $p_{Intervention \times Order} = 0.78$, $p_{Implementation} = 0.63$), as shown in Figure~\ref{fig:total_usage}. One explanation could be that the decreased usage of target apps canceled out with the increased usage of non-target apps, which could include many non-overuse cases.

\begin{figure}[t!]
\centering
\includegraphics[width=\linewidth]{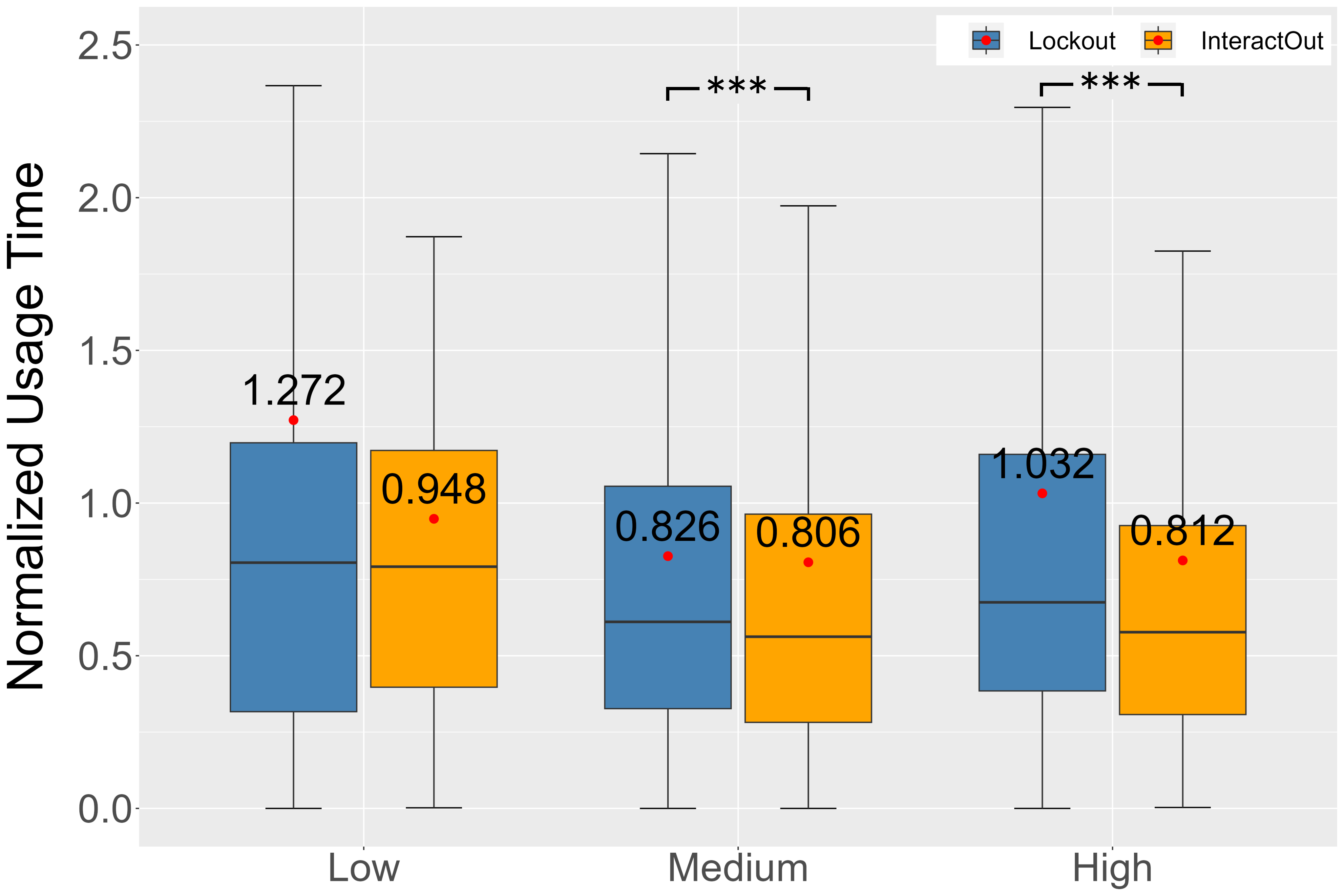}
\caption{{Normalized Target Apps Usage in 3 Levels of Interaction Intensity (Low, Medium, High). Interaction intensity was categorized based on the number of required touch interactions for each target app. It shows the normalized ratio of the usage time against the average usage time in the base week in each interaction intensity level group. Lower ratio means more effective reduction. In each box, the middle black line shows the median and the red point shows the mean. Note that the minimum usage time is 0 as some participants do not use their target apps in some days.}}
\Description{Boxplot of normalized overall usage time in 3 intensity levels: low, meidum and high. The mean normalized usage for the low group is 1.272 for Lockout and 0.948 for InteractOut. The mean normalized usage for the medium group is 0.826 for Lockout and 0.806 for InteractOut. The mean normalized usage for the high group is 1.032 for Lockout and 0.812 for InteractOut.}
\label{fig:target_usage_by_intensity}
\end{figure}

{We further investigated the influence of app interaction intensity on the usage time change. We categorized each target app based on their average number of screen touches per minute (interactions per minute, IPM) during InteractOut's intervention-on period. The thresholds were chosen so that 3 groups had almost the same number of target apps (Low: IPM $\leq 10$: Medium: $10 <$ IPM $\leq 18$; High: IPM $ > 18$). As Figure \ref{fig:target_usage_by_intensity} shows, we found \textbf{both Lockout and InteractOut showed the best performance in the medium-intensity group, while InteractOut showed a significantly stronger usage reduction in the medium and high-intensity group.} A GLMM on fixed effect \textit{Intensity\_Level}, \textit{Intervention} and \textit{Intensity\_Level $\times$ Intervention} showed significance for \textit{Intensity\_Level} ($\chi^2 = 10.5$, $p < 0.05$) and \textit{Intensity\_Level $\times$ Intervention} ($\chi^2 = 6.32$, $p < 0.05$). We then compared Lockout and InteractOut in each intensity group. A GLMM with post-hoc Tukey's HSD test on fixed effect \textit{Intervention} showed that InteractOut had significantly lower usage time than Lockout in the high ($p < 0.001$) and medium ($p < 0.001$) intensity group, but not the low group ($p = 0.36$).
This shows that InteractOut had less impact on users' app usage in low-intensity apps. A potential reason is that the small number of input interactions limits the manipulation of user input.
}

\subsubsection{App Opening Frequency}
\begin{figure*}[t!]
\centering
\begin{subfigure}{.33\linewidth}
\centering
\includegraphics[width=\linewidth]{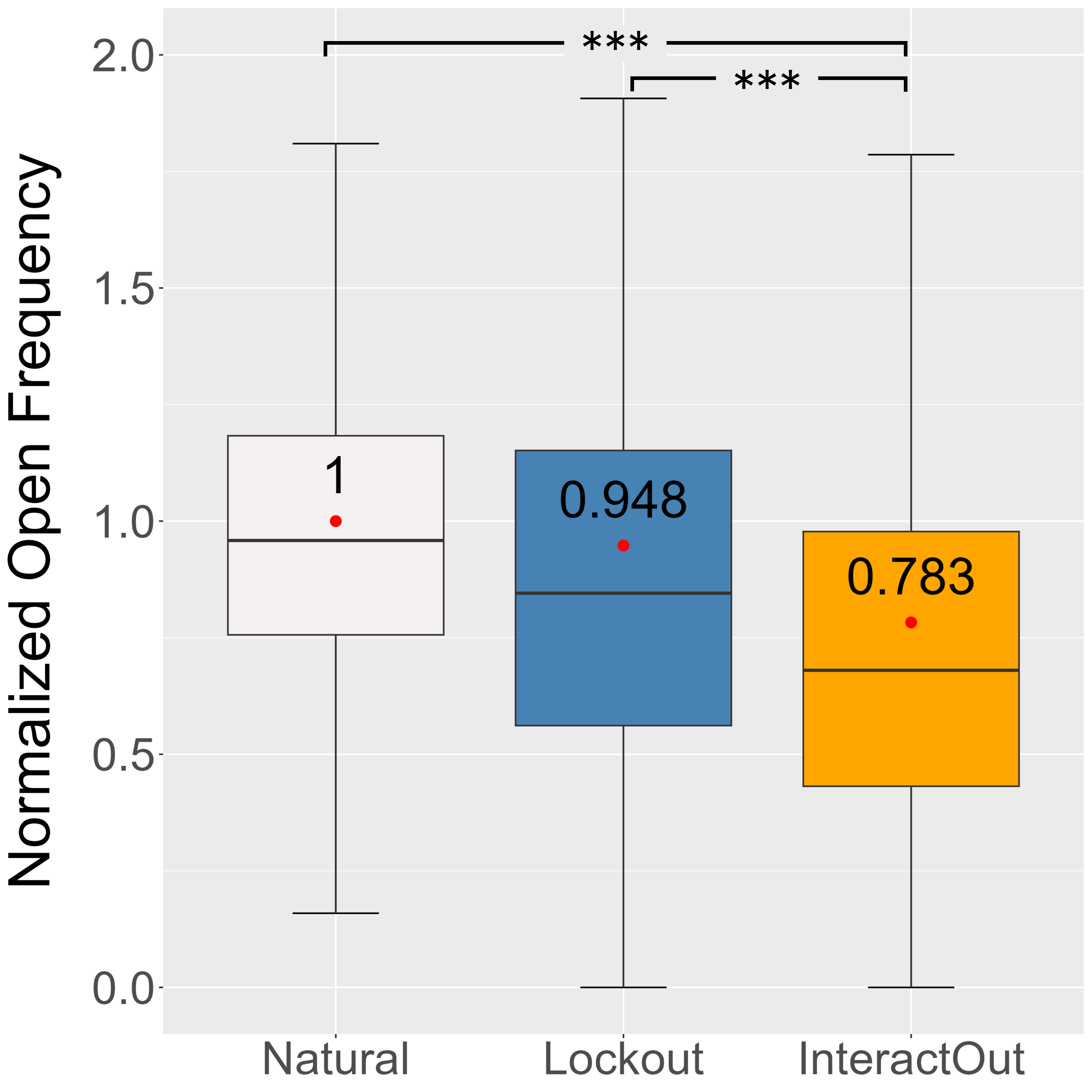}
\caption{Target Apps}
\Description{Boxplot of normalized target app opening frequency. The mean value of Natural (no intervention), Lockout and InteractOut is 1, 0.948 and 0.783. InteractOut is significantly lower than Natural and Lockout, while there is no significance between Lockout and Natural.}
\label{fig:target_freq}
\end{subfigure}%
\hspace{5pc}
\begin{subfigure}{.33\linewidth}
\centering
\includegraphics[width=\linewidth]{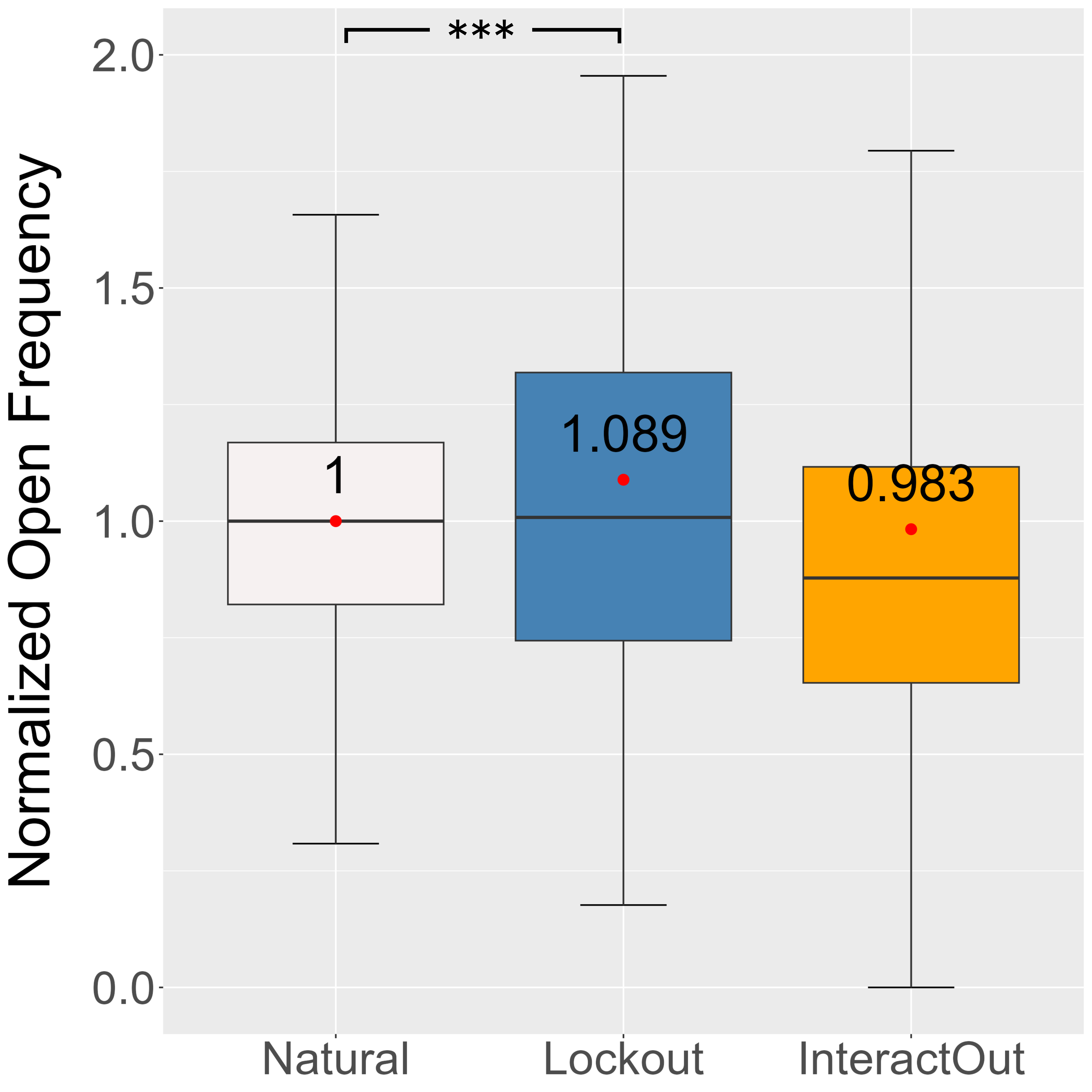}
\caption{Overall Usage}
\Description{Boxplot of normalized overall opening frequency. The mean value of Natural (no intervention), Lockout and InteractOut is 1, 1.089 and 0.983. None of 3 pairwise comparison shows significance.}
\label{fig:total_freq}
\end{subfigure}
\caption{Normalized App Opening Frequency in the Intervention Conditions. It shows the normalized ratio of the usage time against the average usage time in the base week. A lower ratio means more effective reduction. Note that the minimum opening frequency is 0 as some participants do not use their target apps in some days.}
\label{fig:open_freq}
\end{figure*}

\textbf{We observed similar results for opening frequency, where InteractOut reduced smartphone usage significantly more than Lockout.} The average opening frequency was $66$ in the base week and $57$, $48$ for Lockout and InteractOut, respectively.
We also used the normalized data for comparison. Figure \ref{fig:open_freq} shows that participants opened target apps 16.5\% less in the InteractOut period (Mean = 0.78, Std = 0.023) than they did in the Lockout period (Mean = 0.95, Std = 0.024). We ran a GLMM with the same configuration on the normalized opening frequency.
Results showed that \textit{Intervention} was also the main factor of opening frequency changes ($\chi^2 = 56.2$, $p_{Intervention} < 0.001$), while the \textit{Intervention} $\times$ \textit{Order} interaction also had some marginal effect ($p_{Intervention \times Order} = 0.05$, $p_{Order} = 0.84$, $p_{Implementation} = 0.26$). A pairwise post-hoc Tukey's HSD test on \textit{Intervention} showed that InteractOut ($Z = -5.84$, $p < 0.001$) significantly lowered the opening frequency of target apps compared to the base week. In contrast, Lockout showed a similar level of opening frequency as participants' natural behaviors in the base week ($Z = -0.75$, $p = 0.73$). InteractOut showed a significantly lower opening frequency compared to Lockout ($Z = -5.35$, $p < 0.001$).
For overall app usage, a GLMM showed the significance of \textit{Intervention} ($\chi^2 = 20.0, p < 0.001$). The post-hoc test showed significantly higher opening frequency in the Lockout period ($Z = -2.9$, $p < 0.05$) and lower frequency during the InteractOut period, yet without significance ($Z = 1.3$, $p = 0.39$).

\begin{figure}[t!]
\centering
\includegraphics[width=\linewidth]{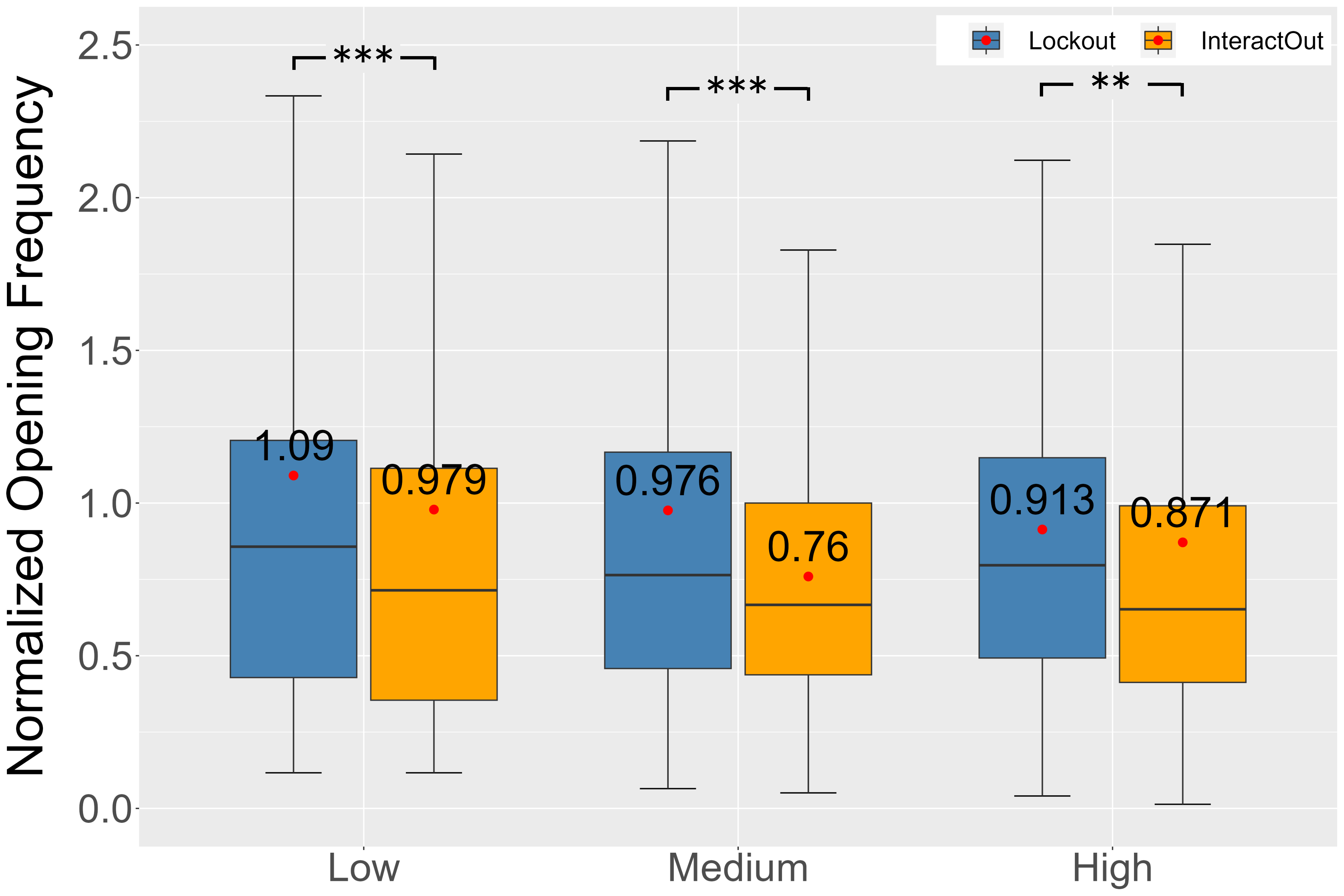}
\caption{{Normalized Target Apps Opening Frequency in 3 Levels of Interaction Intensity (Low, Medium, High). Interaction intensity was categorized based on the number of required touch interactions for each target app. It shows the normalized ratio of the opening frequency against the average opening frequency in the base week in each interaction intensity level group. Lower ratio means more effective reduction. In each box, the middle black line shows the median and the red point shows the mean. Note that the minimum usage time is 0 as some participants do not use their target apps in some days.}}
\Description{Boxplot of normalized overall opening frequency in 3 intensity levels: low, meidum and high. The mean normalized frequency for the low group is 1.09 for Lockout and 0.979 for InteractOut. The mean normalized frequency for the medium group is 0.976 for Lockout and 0.76 for InteractOut. The mean normalized frequency for the high group is 0.87 for Lockout and 0.871 for InteractOut.}
\label{fig:target_freq_by_intensity}
\vspace{-1pc}
\end{figure}

We also investigated the opening frequency change on different app interaction intensities. Figure \ref{fig:target_freq_by_intensity} shows that \textbf{InteractOut had significantly lower opening frequency than Lockout in all 3 intensity groups, but there was no significance between 3 intensity groups.}
A GLMM with post-hoc Tukey's HSD test showed a significantly lower opening frequency in all intensity groups ($p_{low} < 0.001$, $p_{medium} < 0.001$, $p_{high} < 0.01$).

\subsubsection{Lasting Effect on App Behaviors}
\label{subsubsec:lasting_effect}

 \textbf{Both Lockout and InteractOut kept a decreased app opening frequency but did not retain the performance in usage time.} We compared participants' app behaviors on break days to the data on intervention days to investigate the lasting effect of interventions.
A GLMM with the same configuration showed the significance of \textit{Intervention}. A pairwise post-hoc Tukey's HSD test on \textit{Intervention} further showed the decreased opening frequency ($Z = 5.1$, $p < 0.001$) but similar usage time ($Z = 1.6$, $p = 0.27$) in InteractOut break days compared to the base week. Similarly, in Lockout break days, we also found a slightly decreased opening frequency ($Z = 2.9$, $p < 0.01$) and similar usage time ($Z = 1.3$, $p = 0.39$) compared to the base week. There was no sigificant difference between InteractOut and Lockout ($p = 0.95$ for usage time and $p = 0.97$ for opening frequency).

\subsubsection{Results Summary}
Overall, our results on app usage behavior indicated that InteractOut was significantly more effective in reducing target apps' usage time and opening frequency than Lockout. It also showed a lasting effect on the opening frequency of target apps. However, the overall usage across all apps only showed a slight decrease (see Figure~\ref{fig:usage_time} and Figure~\ref{fig:open_freq}). 

\subsection{Intervention Acceptance Rate}

\begin{figure}[t]
\centering
\begin{subfigure}{.8\linewidth}
 \centering
 \includegraphics[width=\linewidth]{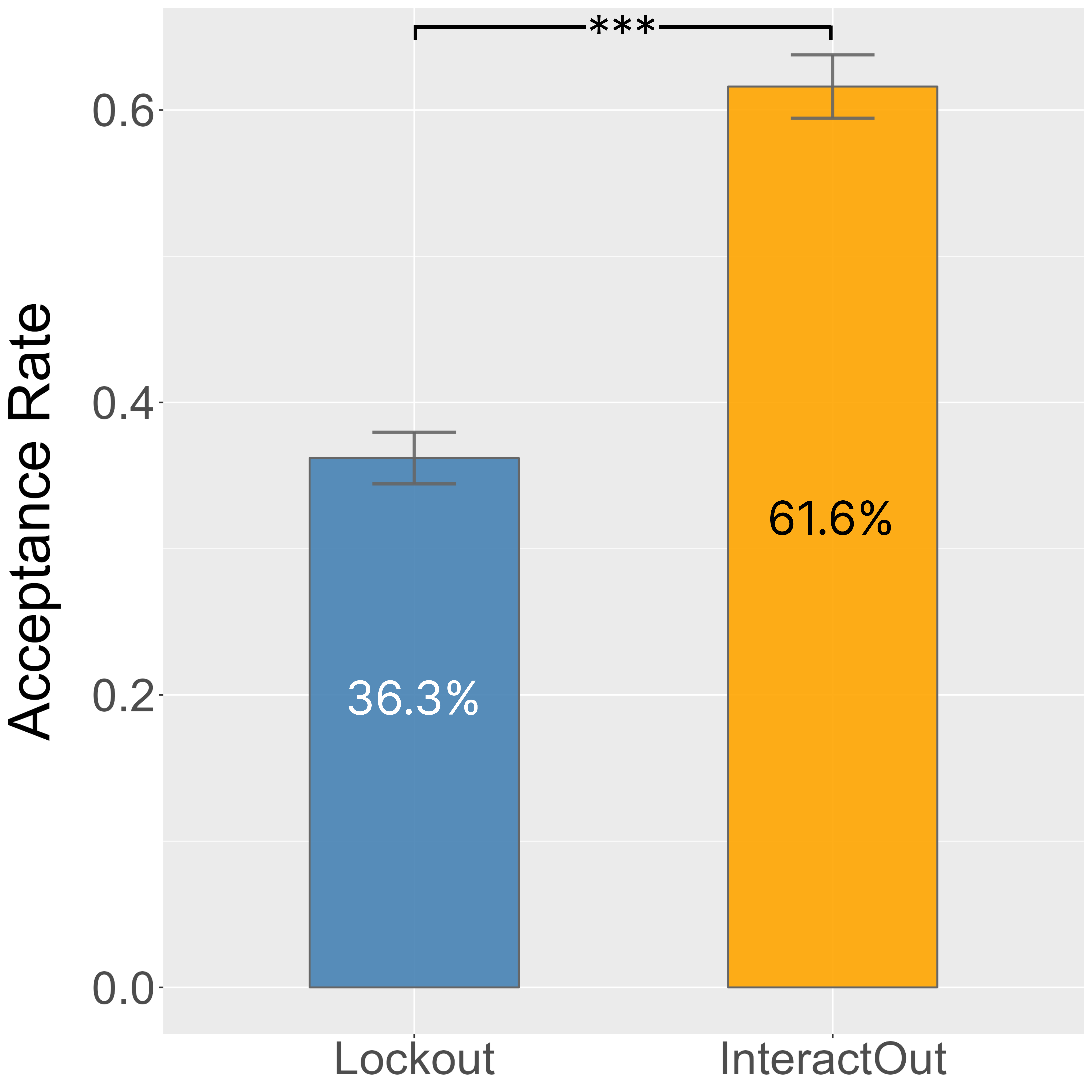}
 \Description{Target apps acceptance rate in 6 app categories. InteractOut has a significantlyu higher acceptance rate in all categories.}
 \caption{Target apps general acceptance rate}
 \label{fig:bypass}
\end{subfigure}
\begin{subfigure}{\linewidth}
 \centering
 \includegraphics[width=\linewidth]{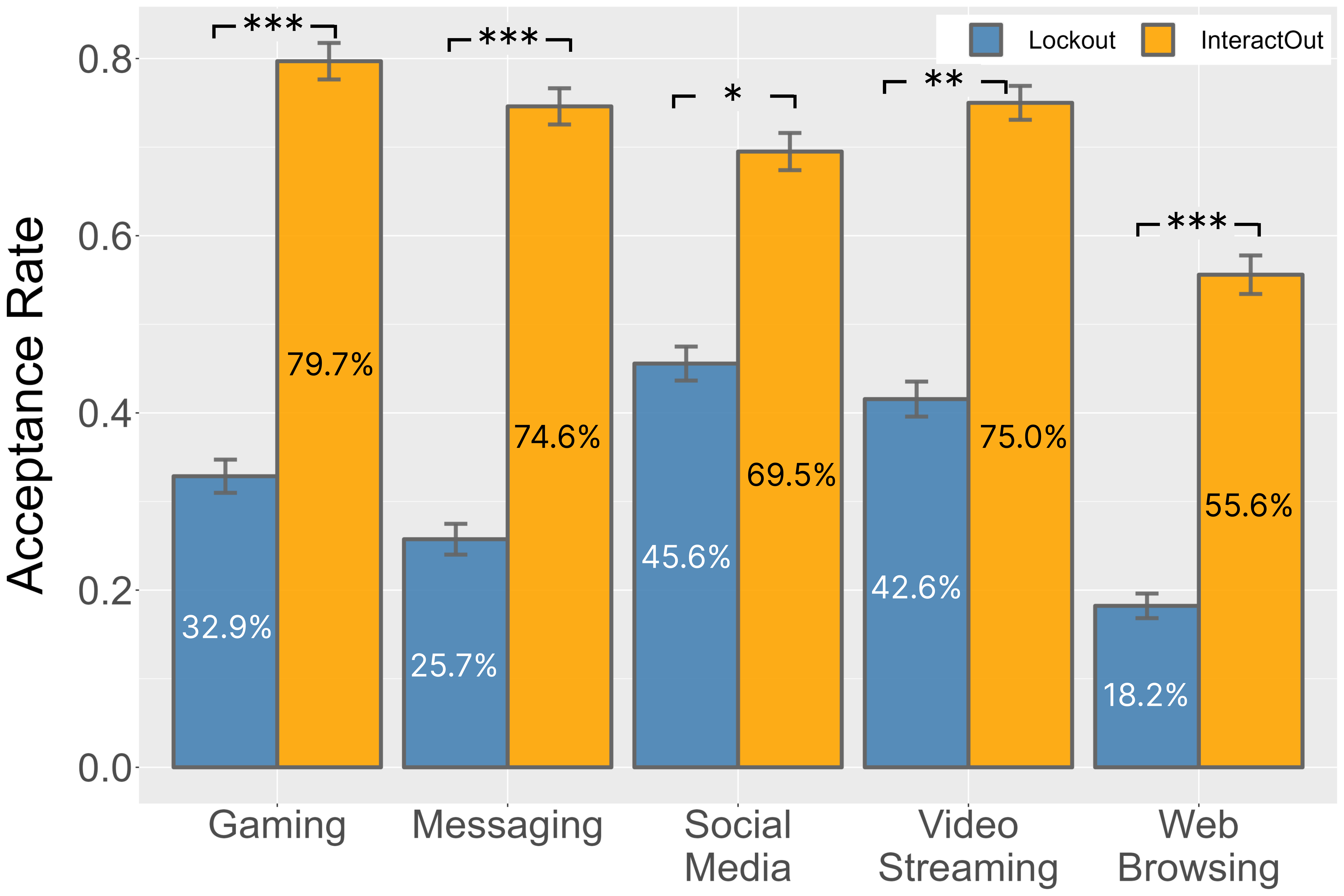}
 \Description{Target apps acceptance rate in 6 app categories. InteractOut has a significantly higher acceptance rate in all categories.}
\caption{Target apps acceptance rate in five app categories}
\label{fig:bypass_app_category}
\end{subfigure}
\caption{Intervention Acceptance Rate, i.e., the proportion of times that the intervention was not bypassed. Higher rate means greater intervention receptivity. InteractOut outperformed Lockout consistently across different app categories.}
\end{figure}

Another important metric of smartphone overuse interventions is their receptivity. We measured the receptivity of the two intervention techniques by their acceptance rate, i.e., the proportion of times participants decided to accept the intervention and did not bypass it. Higher acceptance rate means greater receptivity of the intervention by the users. 
Note that for both InteractOut and Lockout, the bypass options were similar (see Figure~\ref{fig:notification} and Figure~\ref{fig:field_lockout_bypass}).
\textbf{InteractOut showed a significantly higher overall and categorical acceptance rate than Lockout.}
Figure \ref{fig:bypass} shows the overall acceptance rate of the two intervention techniques.
InteractOut had an acceptance rate of 61.6\%, while Lockout was 36.3\%.
A Kolmogorov-Smirnov test found that the acceptance rate followed a Gaussian distribution. Thus, we ran a GLMM with Gaussian link function on fixed effects \textit{Intervention}, \textit{Order}, \textit{App\_Category}, and \textit{Intervention} $\times$ \textit{App\_Category}. Results showed the significance of \textit{Intervention} ($F = 5.61$, $p < 0.05$) and \textit{App\_Category} ($F = 2.80$, $p < 0.05$). The interaction of \textit{Intervention} and \textit{App\_Category} also showed a marginal significance ($p = 0.097$). The post-hoc Tukey's HSD test on \textit{Intervention} showed a significantly higher acceptance rate for InteractOut than Lockout ($Z = 7.6$, $p < 0.001$).

The acceptance rates of each app category are shown in Figure \ref{fig:bypass_app_category}.
A post-hoc Tukey's HSD test on \textit{App\_Category} did not show a statistical difference among app categories. This indicated that InteractOut outperformed Lockout consistently across different app categories.

\subsubsection{Behaviors After Accepting Interventions} 
We also investigated participants' behaviors after encountering the intervention in a target app.
We focused on the immediate events after encoutering an intervention (possible events include entering a non-target app, another target app, or closing the screen), and calculated the proportion of each kind of event.
Our results showed that for InteractOut, participants entered a non-target app 50.8\% of the time, entered another target app 27.1\% of the time, and closed the screen 22.1\% of the time.
In the Lockout condition, entering a non-target app was also the first option (60.1\%), followed by closing the screen (19.8\%) and entering another target app (18.9\%).
Overall, participants had similar behaviors after interventions.

\subsubsection{Results Summary}
Participants showed higher receptivity to InteractOut than Lockout regardless of the categories of their target app choices. But after accepting the intervention, they turned to an intervention-free non-target app most of the time, with a 20\% chance to end smartphone usage. 
This after-acceptance behavior explains the similar usage time and opening frequency increase of overall usage in Section \ref{sec:app_behavior}.

\subsection{Subjective Measurements}
\label{subsec:subjective_measurements}
We also collected participants' subjective feedback through weekly in-study surveys and exit interviews.
Thematic analysis was used to analyze the interview data \cite{braun2006using}, in order to learn about user perceptions of the interventions, including the workload, effectiveness, and receptivity. 
Our results showed that InteractOut was more mentally acceptable than Lockout even though InteractOut introduced a higher cognitive load.

\subsubsection{Subjective Workload}
\begin{figure}[t!]
\centering
\begin{subfigure}{.8\linewidth}
 \centering
 \includegraphics[width=\linewidth]{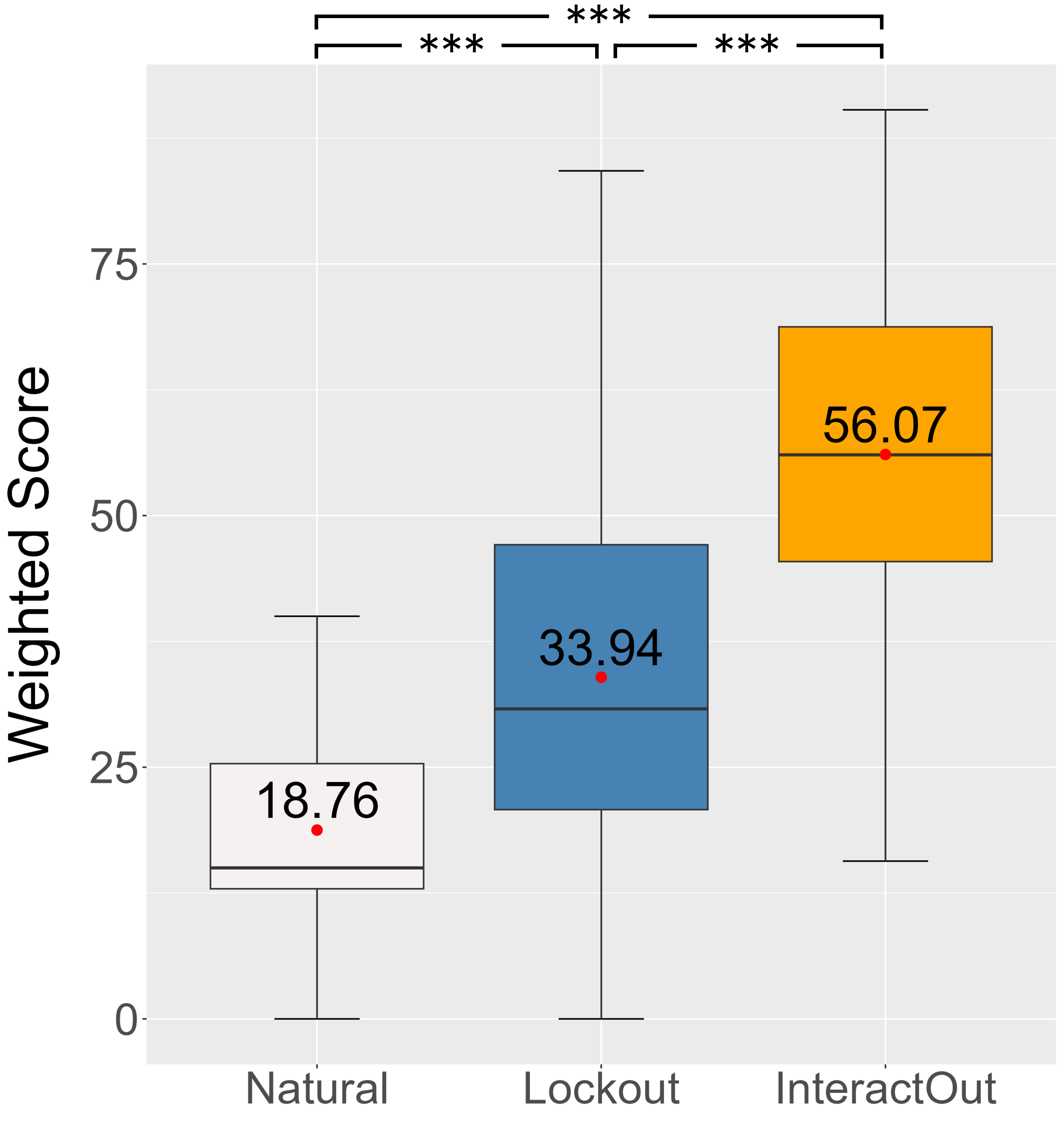}
 \Description{Weighted NASA TLX scores. The mean of Natural, Lockout and InteractOut is 18.76, 33.94 and 56.07. All of 3 pairwise comparisons show significance.}
 \caption{Weighted NASA TLX scores}
 \label{fig:nasa_weighted}
\end{subfigure}
\begin{subfigure}{\linewidth}
 \centering
 \includegraphics[width=\linewidth]{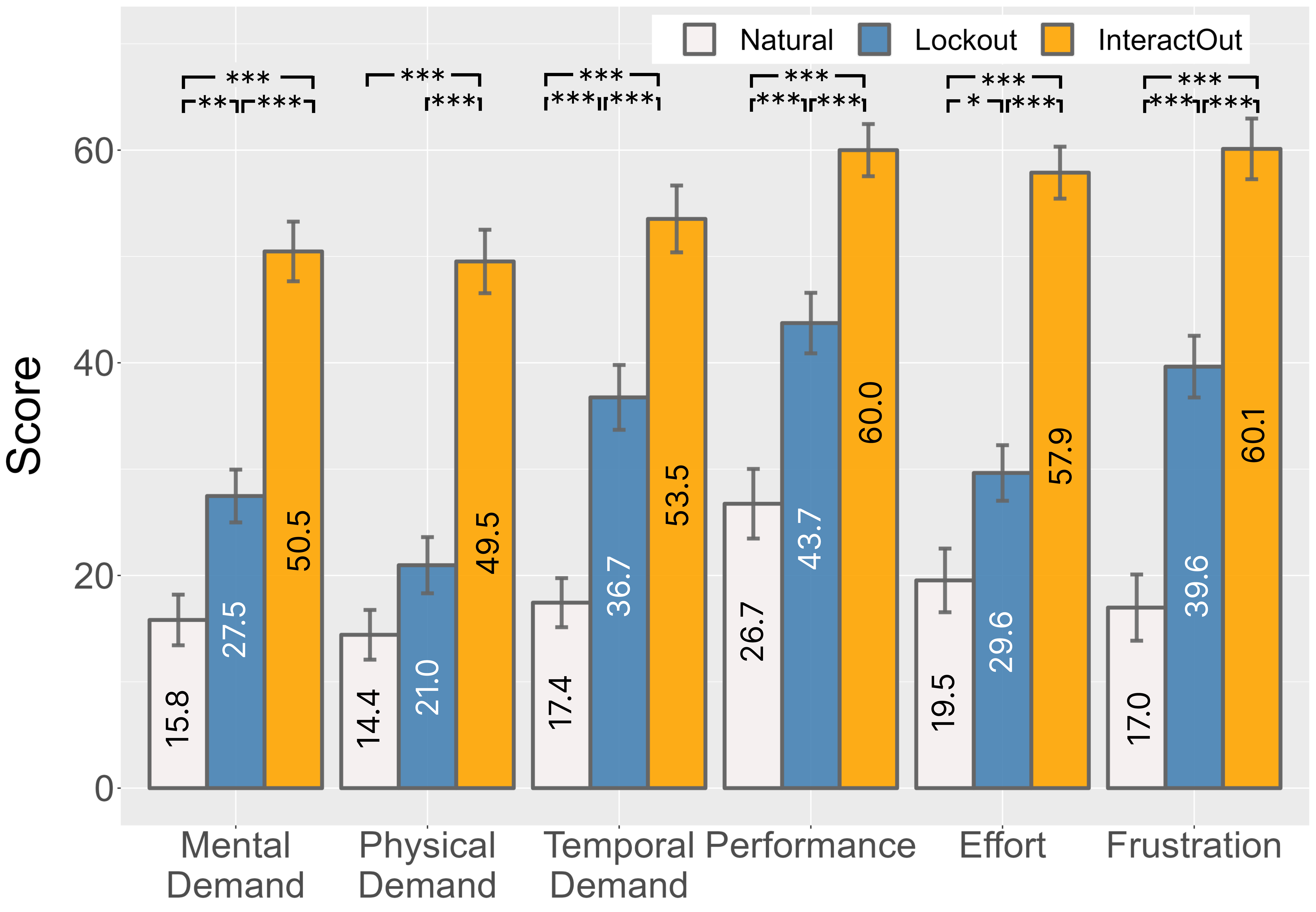}
 \Description{NASA TLX sub-scores. InteractOut has a significantly higher score than both Lockout and Natural in each sub-question.}
\caption{NASA TLX sub-scores}
\label{fig:nasa_per_question}
\end{subfigure}
\caption{NASA TLX Responses in In-Study Surveys. InteractOut has a significantly higher overall and each individual workload.}
\label{fig:nasa}
\end{figure}

We measured the subjective workload of the interventions using the NASA TLX questionnaire in our weekly in-study surveys. We asked participants to score the workload of smartphone use under the current condition (i.e., Natural, InteractOut, or Lockout). We summarize the average overall weighted scores and sub-scores in Figure \ref{fig:nasa}. Expectedly, InteractOut had a higher overall workload than Lockout. A GLMM with Gaussian link function confirmed our observation with the significance on \textit{Intervention} ($\chi^2 = 166, p < 0.0001$) but not others ($p_{Order} = 0.61$, $p_{Intervention \times Order} = 0.64$).

\subsubsection{Perceived Effectiveness and Receptivity}
We summarize the main advantages of InteractOut based on participants' comments.
\begin{s_itemize}
 \item \textbf{Intervene with appropriate flexibility.}
 Most participants appreciated the ``no-force'' intervention InteractOut provided. They were usually in the middle of an activity using the smartphone when encountering the intervention. InteractOut allowed them to finish what they were doing but also acted as a gentle reminder.
 For example, \textit{``[InteractOut] is more comfortable because it allows me to complete my current task such as reading a news article.''} (P26) \textit{``InteractOut is more gentle. I do not have to snooze. It is more user-friendly and constantly reminds me to put down my phone.''} (P35)
 Participants generally found such soft and implicit interventions to be more acceptable.
 \item \textbf{Provide user agency.}
 In traditional lockout interventions, users often have no choice but to stop or bypass the limit. Even if they choose to stop, they may feel little agency in making this decision. However, with InteractOut, participants could decide when to stop, which is more natural than a restrictive lockout. {This enabled them to finish their current task and have a graceful stop.} \textit{``It slows down the app, and I start to think if I want to use the app anymore, then I feel like I do not need it and I will stop.''} (P25) {\textit{``(InteractOut) is better as the overlay at least allows me to finish reading the article.''} (P3)} The decision-making process was controlled by the users themselves and presented continuously, thus leading to better effectiveness and user experience. 
 \item \textbf{Be always-on.}
 Participants experienced the intervention consistently once they exceeded the time limit. Being always on can act as a constant reminder for users. \textit{``It is constantly there and I am always aware of it.''} (P33) Such a user experience change could sometimes be ``misperceived'' as a change on the app itself. \textit{``When I use the app [in InteractOut], I found I do not like using the app, whereas the blocking [Lockout] does not change how the app works after I ignore the limit.''} (P21) In this case, participants' mental model towards the app changed due to constant intervention.
 \item \textbf{Introduce a gulf of gratification.}
 Although InteractOut's intervention is implicit, it could introduce a gulf of gratification, meaning users needed to spend more effort to gain entertainment from smartphones.
 Participants reported that InteractOut introduced gaps in simple operation logic such as scrolling up (e.g., when using social media or video streaming apps).
 \textit{``[InteractOut] slowing down the scroll speed, [and]... The scrolling is also not real-time... I lose interest of the app.''} (P39)
 This indicates that manipulating input can effectively trigger System 2 in smartphone overuse scenarios~\cite{alter2007overcoming,taleb2014antifragile}.
\end{s_itemize}

\subsubsection{Critiques}
We also received some negative feedback about InteractOut. First, InteractOut made some precise and real-time operations difficult. This was important for non-casual games, such as role-play, first-person shooting, and racing games. For example, P27 picked two games for intervention, but could not play the game with InteractOut.
Second, InteractOut was not as effective at intervening in scenarios with sparse interactions, such as watching a long video or reading an e-book. \textit{``In Netflix, there is not too much operation, I cannot sense the intervention.''} (P33) In these cases, participants spent most of the time looking at the screen without performing many gestures, making InteractOut less effective. However, we envision that InteractOut could be combined with other existing intervention techniques. We will discuss some potential cases in Section~\ref{subsec:discussion_existing_intervention}.

\subsubsection{Mismatch between Real \& Perceived Effectiveness}
\label{subsubsec:mismatch}
Moreover, we found an interesting mismatch between the real and perceived effectiveness of interventions for some participants.
For example, P31 considered InteractOut as ineffective, but the data showed that their usage time on target apps decreased by 17.9\% (from 373.1 minutes to 306.1 minutes). A similar mismatch was also observed in four other participants.
This suggests that there may exist a subtle and implicit mental effect of input manipulation interventions. Users may not feel the effectiveness explicitly, yet their objective phone usage behavior has changed.
However, the sample of 5 participants for this observation is too small for an in-depth investigation.
Future work with a larger-scale deployment could enable us to further investigate this, which we discuss in Section~\ref{subsec:discussion_future_deployment}.

\subsubsection{Subjective Results Summary}
InteractOut was perceived as being effective and acceptable for most participants due to its subtle and continuous manipulation of input. 
This property shows a significant advantage over traditional disruptive and restrictive interventions.
Users are more likely to develop an active control of reducing smartphone usage, which is one of the favorable and ultimate goals of smartphone overuse intervention.
Meanwhile, we acknowledge the disadvantage of InteractOut for scenarios where smartphone usage is more consumption-focused without heavy input interactions.
\section{Discussion and Future Work}
In this work, we propose a set of new input manipulation-based intervention techniques InteractOut. Our lab study showed the effectiveness of these techniques, while highlighting four promising interventions. This pointed us to our field experiment to further evaluate these four interventions' performance in real-world scenarios.
Through our lab study and field experiment, we showed the advantage of InteractOut.
In this section, we discuss the results from our study, the potential combination of InteractOut and existing intervention methods, the implications for various stakeholders, as well as future directions for larger-scale real-life deployments. We also discuss the limitations of our work.

\subsection{Workload and Subjective Preference}
As shown in Section \ref{subsec:subjective_measurements}, InteractOut introduced a higher perceived workload for users when operating their smartphones. However, the majority of participants favored InteractOut as their daily smartphone intervention. The primary reason was that InteractOut allowed them to maintain a desirable level of flexibility and control over their smartphone usage. Despite the increased workload, participants were willing to bear additional interaction costs to maintain control. This result aligns with prior research emphasizing the limitations of forceful interventions, which often lead to user abandonment~\cite{CommericalAppComp}.

This finding also shows that InteractOut strikes a delicate balance between intervention effectiveness and user acceptance. However, the current workload of InteractOut may still not be low enough, which require future investigations. The challenge lies in determining the optimal balance at which workload serves as a constructive motivator for behavior change without overwhelming users. This balance between intervention subtlety and user engagement needs further exploration in the development of future smartphone overuse interventions.

\subsection{{Comparison with Existing Interventions}}

{Although we did not formally compare with additional baselines in our field study, here we compare InteractOut with some existing works using the respective statistics reported in the results sections. Compared to the respective baseline usage without interventions, InteractOut had a higher reduction (-21.9\%) in usage time than LocknType \cite{LocknType} (-7.8\%), Good Vibration \cite{Goodvibrations} (-15.7\%), and TypeOut \cite{TypeOut} (-20.3\%). InteractOut also had a higher acceptance rate (61.3\%) than TypeOut \cite{TypeOut} (58.3\%). On the other hand, InteractOut had a lower reduction in opening frequency (-21.7\%) than TypeOut \cite{TypeOut} (-27.8\%). These preliminary comparisons show the potential of InteractOut to have a promising performance compared to prior works, while further study is needed to provide quantitative evidence, and it is difficult to directly compare their results because of the different study populations, durations, and other factors. Future work could also devise benchmarks and guidelines \cite{NEURIPS2022_9c7e8a08,xu2023globem} to enable such comparisons across research studies, which is often difficult for human-subject experiments.
}

\subsection{Compatibility and Complementarity with Existing Interventions}
\label{subsec:discussion_existing_intervention}
In considering the potential applications of InteractOut within the landscape of smartphone overuse interventions, we foresee that it could offer compatibility and complementarity to both explicit and implicit intervention techniques.
InteractOut could be seamlessly integrated with existing explicit interventions, such as Timed Lockout. This integration would involve applying InteractOut during the extra time granted after users have reached their predefined smartphone usage limit.

Furthermore, in the realm of implicit interventions, InteractOut exhibits compatibility with reminder-based techniques. This synergy allows for a multi-pronged approach to combining InteractOut's subtle input manipulation and periodic reminders, both aimed at redirecting users' attention away from their devices.
Additionally, InteractOut's adaptability extends to other implicit interventions, such as Good Vibrations \cite{Goodvibrations}, or potential user interface and user experience manipulations designed to gently discourage excessive smartphone usage.

Since InteractOut mainly focuses on the input aspect, it could serve as a valuable addition to other intervention strategies. Its compatibility with existing techniques and potential for synergy offer researchers, developers, and users a promising pathway to more effectively manage smartphone usage, while maintaining a user-centered approach to intervention design.

\subsection{Implications for Users, Platform Owners and App Developers}

As highlighted in Section \ref{subsec:subjective_measurements}, some participants perceived InteractOut as an alteration of their usual app interactions. {On the one hand, it raises concerns about the potential emergence of a dark pattern \cite{gray2018dark}. The implicit manipulation of user input could be misused and considered deceptive. Users may perceive this as a violation of their autonomy and control, potentially leading to frustration and mistrust. We also recognize the possibility that users could mis-recognize the intervention as device malfunction. A potential solution is to provide a small indicator, such as a notification or a visual indicator to remind users of the consented intervention. On the other hand,} this implicit nature of the intervention could also pose challenges for platform owners like iOS and Android, particularly considering the constraints imposed on third-party apps accessing accessibility services by platforms like Google \cite{vonau_2022_android}. Platform owners and app developers, who often prioritize fine-tuning the user experience to maximize usability and engagement, may also have reservations about interventions that modify their the behavior of their systems and apps. InteractOut, in its current form, functions as an external third-party tool, sidestepping the need for explicit buy-in from platform owners and app developers.

However, an avenue for future research lies in investigating the perceptions and reactions of these stakeholders towards interventions like InteractOut. Understanding their perspectives could inform refinements to make the intervention more compatible with platform guidelines and potentially gain direct support within individual apps. Balancing the goals of reducing smartphone overuse with the interests of platform owners and app developers presents an intriguing challenge for the evolution of interventions like InteractOut. Further explorations in this direction could pave the way for more harmonious integration into the broader digital ecosystem.

\vspace{-1pc}
\subsection{Large-Scale Longitudinal Deployments \& Future Directions}
\label{subsec:discussion_future_deployment}
Both our lab study and field experiment only involved a limited group of users (30 and 42, all university students), with a limited period of time for deployment.
Here, we list a few promising directions when large-scale longitudinal deployment is possible.

\textbf{User-Defined Intervention Configuration and Personalized InteractOut}.
In our experiment, we pre-determined the time budget and intervention combinations for the purpose of the study. But in real-life deployment, users should have the agency to control the intervention.
For example, our config interface (similar to the one shown in Figure~\ref{fig:config_page}) could be extended to enable users to pick their own preferred combination. They could also set their own intensity and the way intensity changes over time. Such a customized configuration could better fit individuals' preferences. Moreover, future work could investigate the potential of making interventions intelligent and adaptive based on AI models to detect risk behavior with sensor data (e.g., \cite{xu2019leveraging}), moving toward the vision of just-in-time adaptive intervention (JITAI)~\cite{nahum2018just,orzikulova2024time2stop}.

\textbf{App Feature-Level Interventions}. Our current version of InteractOut mainly focuses on intervening at the system-level. It is triggered in a specific set of target apps, but the intervention remains the same across different apps. This sometimes can cause issues, especially when an app is complicated, and not all functions need to be intervened. A beneficial intervention should support meaningful use (i.e. messaging) and limit meaningless use (i.e. endless news/posts feed)~\cite{CommericalAppComp}.
We also believe that the real-perceived effectiveness mismatch mentioned in \ref{subsubsec:mismatch} could be addressed by adding app feature-level support to InteractOut.
In larger-scale deployments, we could extend our exploration to certain apps' specific feature-level interventions~\cite{orzikulova2023finerme}.
For example, InteractOut may only intervene in the posting and browsing features of a social media app, but not influence its messaging function, since the latter actively involves social communication. 
We aim to explore how such interventions may be harnessed to augment user receptivity and enhance the overall effectiveness of the intervention strategy.

\textbf{Input Manipulation beyond Smartphones}.
Unlike traditional interventions on the screen output, InteractOut manipulates user inputs to create interventions.
Although InteractOut is implemented on smartphones, this idea could be generalized to other devices beyond smartphones with different input modalities.
For example, on desktops with keyboard and mouse control, we could design similar delay interventions on mouse pointer movement, double clicking interventions on mouse clicking, and key remapping interventions on the keyboard. On tablets with a mixture of smartphone and desktop operation logic, we could apply either gesture or keyboard interventions when appropriate. 
We also envision the potential of input manipulations for future devices such as augmented reality and virtual reality (AR/VR).
Our InteractOut design space has the potential to be applied and extended to various scenarios to generate more input interventions.

These future research directions with large-scale deployments not only enrich our understanding of the multifaceted potential of InteractOut but also contribute to the ongoing evolution of input manipulation-based intervention methodologies.

\subsection{Limitations}
There are also some limitations in our work.
First, our study population mainly focuses on college students in a local university. The results may not be representative or generalizable to other population groups.
Moreover, in the current implementation, participants needed to release their fingers to receive feedback. For swipe interactions, this means that participants could only see the effect of the swipe after the completion of the trajectory. For long swipes, this would cause a recognizable delay.
This problem could potentially be resolved in the MotionEvent level~\cite{motionevent}, which act on each motion event and preserve the fluency of the whole gesture. However, by the time we wrote this paper, we had yet to find an API to provide this fine-grained manipulation. To enable this, Android could provide new APIs to developers to manipulate the global motion events, or incorporate a hardware-level implementation of InteractOut that can directly control feedback of the touchscreen.
\section{Conclusion}

In this paper, we proposed a new perspective on smartphone overuse intervention. We designed a suite of input manipulation-based intervention techniques called InteractOut. 
Through a pilot lab study we identified four promising InteractOut implementations and their appropriate intensity settings. We then conducted a 5-week field experiment with a traditional Timed Lockout baseline. InteractOut showed 15.6\% lower usage time, 16.5\% lower open frequency, and 25.3\% higher acceptance rate on target apps with a better receptivity.
InteractOut suggested the long-term benefit of forceless intervention with a precisely controlled user experience degradation, which should be a joint effort between platform owners, app developers and intervention designers.
We envision InteractOut to supplement current smartphone interventions and inspire a wide range of input-based intervention techniques for interactive devices. 

\begin{acks}
We thank our participants for their contribution to our lab and field experiments, and the reviewers for their valuable feedback and suggestions. This research is supported in part by a Google Cloud Platform Credit Award.
\end{acks}

\bibliographystyle{ACM-Reference-Format}
\bibliography{reference}

\end{document}